\documentclass[aps,prl,twocolumn,showpacs]{revtex4-1}
\usepackage{epsfig, amsmath, amsfonts, amssymb, bm, graphicx}
\usepackage{multirow}
\usepackage{rotating}
\usepackage{threeparttable}

\begin{document}

\DeclareGraphicsExtensions{.eps,.EPS,.jpg,.pdf}

\title{Cooling All External Degrees of Freedom of Optically Trapped Chromium Atoms Using Gray Molasses}

\author{L. Gabardos, S. Lepoutre, O. Gorceix, L. Vernac, B. Laburthe-Tolra}
\affiliation{1 Universit\'e Paris 13, Sorbonne Paris Cit\'e, Laboratoire de Physique des Lasers, F-93430
Villetaneuse, France and 2 CNRS, UMR 7538, LPL, F-93430 Villetaneuse, France}

\date{\today}

\begin{abstract}
We report on a scheme to cool and compress trapped clouds of highly magnetic $\mathrm{^{52}Cr}$ atoms. This scheme combines sequences of gray molasses, which freeze the velocity distribution, and free evolutions in the (close to) harmonic trap, which periodically exchange the spatial and velocity degrees of freedom. Taken together, the successive gray molasses pulses cool all external degrees of freedom, which leads to an increase of the phase-space density ($P.S.D.$) by a factor of $\approx250$, allowing to reach a high final $P.S.D.$ of $\approx1.7^{-3}$. These experiments are performed within an optical dipole trap, in which gray molasses work equally well as in free space. The obtained samples are then an ideal starting point for the evaporation stage aiming at the quantum regime.
\end{abstract}



\maketitle

\section{Introduction}

To date, three highly magnetic atomic elements have been cooled down to degeneracy: chromium ($S=3$) \cite{Griesmaier2005,Beaufils2008}, dysprosium ($J=8$) \cite{Lu2011} and erbium ($J=6$) \cite{Aikawa2012}. The quantum gases made of these elements display several unusual properties induced by the importance of the dipole-dipole interaction between the constituent atoms \cite{Lahaye2009,Baranov2008}. Because this interaction is long-range and anisotropic, it qualitatively differs from the contact interaction, which usually dominates the physics of atomic quantum gases. It is difficult to pay tribute to all the groundbreaking results already reported but, to cite a few, let us mention: the spectacular $d-$wave-like instability \cite{Lahaye2008} and the evidence for intersite dipolar interaction in a Cr-BEC \cite{DePaz2013} (respectively in bulk and in an optical lattice); the observation of the roton mode in a Er-BEC \cite{Chomaz2018} and that of quantum droplets in a Dy-BEC \cite{Schmitt2016}. Cr atoms carry a smaller permanent magnetic dipole than Dy and Er atoms; however they have the specificity of being spin-3 atoms in their ground state, allowing Cr quantum gases to remain prominent systems for the study of the magnetic dipole-dipole interaction within quantum gases and enabling the exploration of spinor physics from a new perspective \cite{Pasquiou2012,Naylor2016}.

Since the early attempts, laser cooling of Cr atoms has been hampered by high light-assisted losses within magneto-optical traps (MOTs) \cite{Bradley2000,Chicireanu2006}. As a consequence, various procedures had to be developed to reach degeneracy \cite{Schmidt2003,Beaufils2008}, resulting in Cr quantum gases with relatively low atom numbers after forced evaporative cooling is performed \cite{Griesmaier2005,Beaufils2008}. Increasing the number of atoms available at the beginning of evaporation would thus be highly desirable in order to improve these counts.
Here, we demonstrate laser cooling of a Cr ensemble using 3D gray molasses, following the approach first demonstrated with Cs atoms in the 1990's \cite{Boiron1995}, and extended with spectacular results in the 2010's first to $\mathrm{^{39}K}$ and $\mathrm{^{41}K}$ \cite{Landini2011} and then to $\mathrm{^{40}K}$ \cite{Fernandes2012}, $\mathrm{^{7}Li}$ \cite{Grier2013}, $\mathrm{^{6}Li}$ \cite{Burchianti2014}, $\mathrm{^{23}Na}$ \cite{Colzi2016}, $\mathrm{^{87}Rb}$ \cite{Rosi2018} and metastable $\mathrm{^{4}He}$ \cite{Bouton2015}. First evidence for 1D-cooling of a chromium thermal beam using gray molasses in the $lin\perp lin$ configuration was reported in \cite{Drewsen1996} but we are not aware of any more recent experiments along this line.

\begin{figure}[]
	\centering
	\includegraphics[width=0.9\columnwidth]{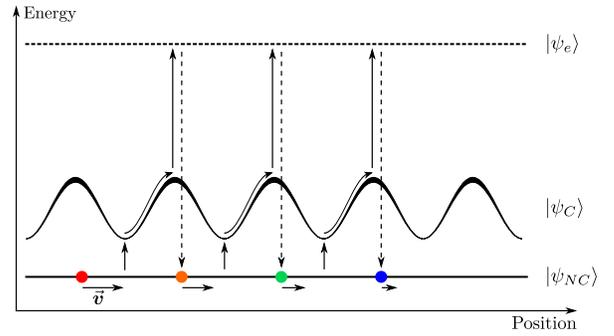}
	\caption{Principle of the Sisyphus cooling and VSCPT taking place during the GM stage. Interference between the different GM laser beams leads to a sinusoidal periodic AC Stark shift potential for the coupled state $\left|\psi_C\right>$. However, the non-coupled state $\left|\psi_{NC}\right>$ does not see the light and therefore has a flat potential. Transitions due to the laser light (long solid arrows) between $\left|\psi_C\right>$ and the excited state $\left|\psi_e\right>$ occur preferentially in the upper part of the potential hills as illustrated by a thicker aspect of the periodic potential. There, atoms are rapidly optically pumped toward $\left|\psi_{NC}\right>$ (dashed arrows). Transitions due to atomic motion (short solid arrows) from  $\left|\psi_{NC}\right>$ to $\left|\psi_C\right>$ occur preferentially in the lower part of the potential hills.}
	\label{molasses}
\end{figure}

We find that the gray molasses approach is an excellent means to circumvent the laser-cooling difficulties in chromium. The main reason is that laser-cooling with gray molasses in the $\sigma^+-\sigma^-$ configuration uses a combination of velocity selective coherent population trapping (VSCPT) and Sisyphus cooling due to the presence of different kinds of light polarization and intensity gradients \cite{Hopkins1997}. The VSCPT insures that atoms with small velocities remain in states not coupled to the light (the "dark states") built by the dressing induced by the gray molasses (GM) laser beams for any $J\rightarrow J$ or $J\rightarrow J-1$ transition. Moving atoms in the coupled states lose kinetic energy through Sisyphus cooling while moving atoms in the non-coupled states can return to the coupled state thanks to motional couplings (Fig.~\ref{molasses}); these non-adiabatic transitions occur with a $v^2$ dependence \cite{Weidemuller1994} such that the coldest atoms are stored in dark states where they are protected from light-assisted collisions. The Sisyphus cooling mechanism benefits from the high friction force even more so because chromium involves blue ($\lambda=429$ nm) cooling light and the friction coefficient scales as $1/\lambda^2$ \cite{Dalibard1989}.

\begin{figure}[]
	\centering
	\includegraphics[width=0.9\columnwidth]{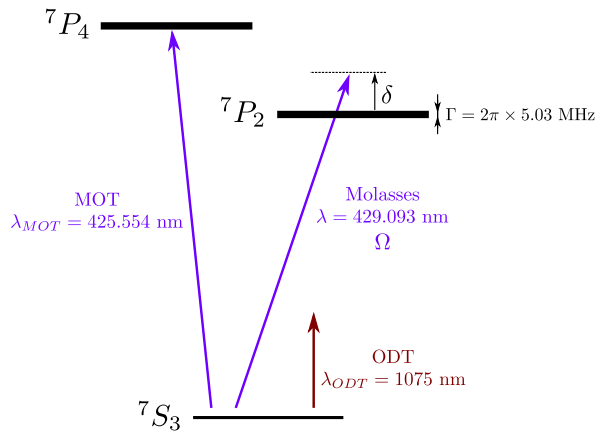}
	\caption{Relevant energy levels and transitions used for the magneto-optical trap, the optical dipole trap (ODT) and gray molasses.}
	\label{levels}
\end{figure}

Furthermore, the absence of hyperfine structure in the presently studied case of bosonic Cr results in a simple scheme with no need for repumpers. Our laser beam arrangement for implementing the gray molasses therefore involves only three retro-reflected laser beams in the $\sigma^+ - \sigma^-$ configuration along three orthogonal directions of space, with the laser frequency blue-detuned with respect to the ${^7S_{3}} \rightarrow {^7P_{2}}$ transition at $429$ nm (Fig.~\ref{levels}).

We find that our gray molasses cooling scheme brings all atoms from a pre-cooled temperature of $60$ $\mu$K to $8$ $\mu$K in $3.5$ ms, thus insuring that light-assisted losses are negligible \cite{Bradley2000,Chicireanu2006}. The temperature which is reached $T \approx 4\ T_{rec}$ (where $T_{rec}\approx 2$ $\mu$K is the recoil temperature), is close to the ultimate limit for gray molasses beyond which the capture velocity gets below the \emph{r.m.s.} velocity of the atoms in the thermal cloud \cite{Dalibard1989}.

\begin{figure}[]
	\centering
	\includegraphics[width=0.9\columnwidth]{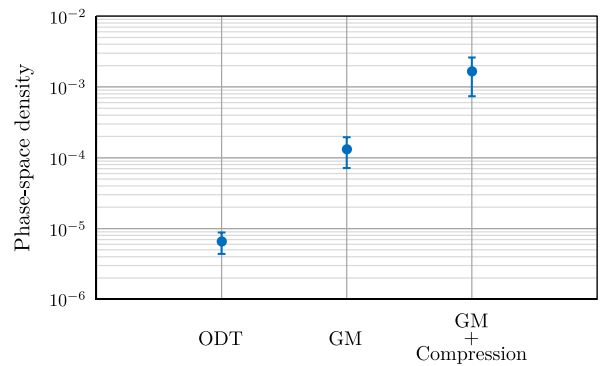}
	\caption{Evolution of the phase-space density when atoms are loaded in the optical dipole trap, when a single gray molasses pulse is performed (GM) and when the scheme to cool all external degrees of freedom with multiple GM pulses is applied.}
	\label{PSD}
\end{figure}

Additionally, we find that Cr gray molasses work equally well in free space as within an optical dipole trap as already reported for cesium \cite{Boiron1998} and lithium atoms \cite{Burchianti2014}. This allows us to perform successive pulses of gray molasses inside the trap, separated by a free evolution time of the atomic cloud. For our parameters, the trapping potential is nearly harmonic. A free evolution in the trap therefore results in a $\pi/2$ rotation in phase-space when the evolution time is a quarter of the trap period $T/4$ (thus exchanging spatial and velocity degrees of freedom). Two molasses pulses performed in the optical dipole trap (ODT) separated by a $T/4$-long free evolution can therefore cool both the spatial and velocity distributions. This technique was first applied in \cite{DePue1999,Hu2017} (although not in the context of gray molasses) and the phase-space manipulation is also reminiscent of the delta-kick cooling technique \cite{Ammann1997,Marechal1999}. The combination of cooling and compression techniques applied along the three space axes insures that all external degrees of freedom are cooled. We reach a temperature below $10$ $\mu$K in the optical trap. The overall increase in the phase space density brought by this procedure is between two and three orders of magnitude, leading to a trapped sample with $\approx 4.10^5$ Cr atoms at a very favorable phase-space density of $(1.7\pm 0.9)\times 10^{-3}$ (Fig.~\ref{PSD}).

\section{gray molasses, main mechanisms}

\begin{figure*}[]
	\centering
	\includegraphics[width=0.9\textwidth]{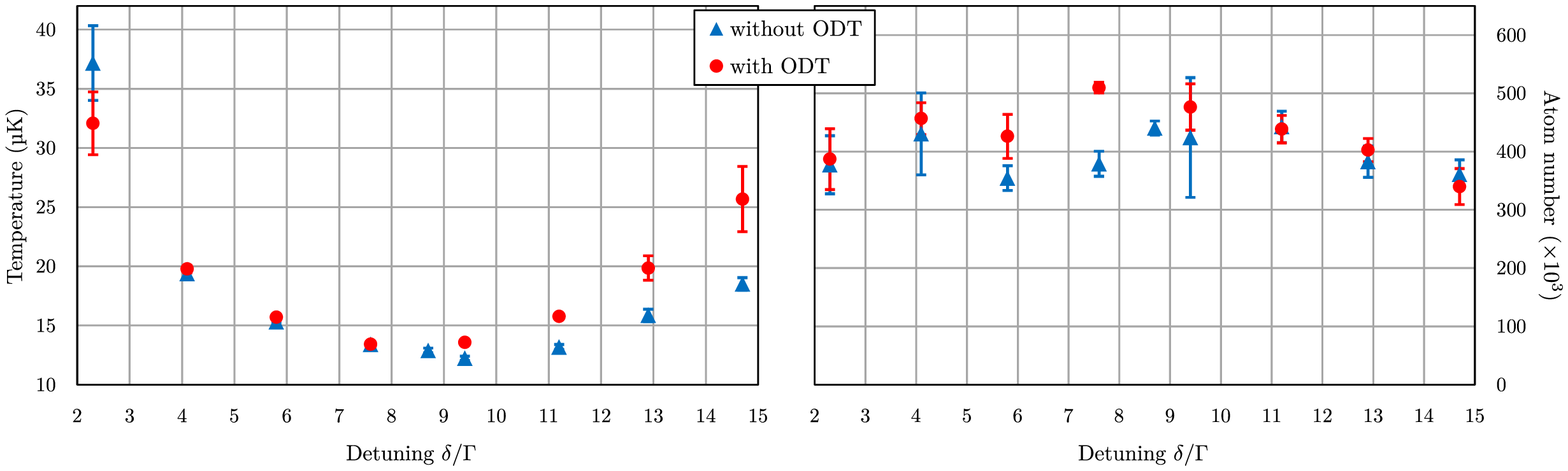}
	\caption{Evolution of the temperature (left) and the atom number (right) with respect to the detuning from the $^7S_3 \rightarrow {^7}P_2$ transition with the optical dipolar trap turned on (red dots) or off (blue triangles). In all the graphs in this paper, the error bars indicate $\pm\sigma$ with $\sigma=\sqrt{\frac{1}{N}\sum_{i=1}^{N}(x_i-\overline{x})^2}$ the standard deviation obtained from $N$ data points (typically $N\approx5$).}
	\label{graphes1}
\end{figure*}

We now introduce the main physical mechanisms involved in gray molasses. First, the use of a $J \rightarrow J-1$ transition insures that there exists at least one dark state, formed by a superposition of different Zeeman states in the electronic ground state. The general form of those dark states can be written $\psi_{DS}=\sum_{m_S}\alpha^{DS}_{m_S}\left|m_S\right>$. The actual linear superposition $\alpha^{DS}_{m_S}$ depends on the Rabi frequencies, the detunings, and the phases of the different laser beams. For this reason, the coefficients $\alpha^{DS}_{m_S}$ indirectly depend on the velocity $v$ of the atoms. In other so called bright dressed states, absorption of light is possible then followed rapidly by spontaneous emission.

When atoms are in the bright states, they experience a position dependent periodic AC Stark shift associated with the molasses beams. This periodic light-shift arises from the polarization gradients created by the laser beams. Atoms also experience a corresponding spatial dependence in the dissipative part of their interaction with light. Crucially, blue detuning the laser frequency compared to resonance insures that dissipation is highest when the potential of the AC Stark shift is maximum. This results in a Sisyphus mechanism such that optical pumping is mostly happening where potential energy is maximum \cite{Dalibard1989}.

Blue detuned molasses in a $J \rightarrow J-1$ configuration are particularly efficient because optical pumping populates dark states. If the velocity of atoms is close to zero, atoms can remain in these dark states for very long times, in a VSCPT scenario. If however atoms possess a finite velocity, their motion through the polarization gradients results in a dynamical change of the dark dressed states (\emph{i.e.} of the $\alpha^{DS}_{m_S}$). This leads to a non-adiabatic coupling from the dark states to the bright states with a rate scaling as $v^2$ and occurring preferentially when the energies of the dark and bright states are closest due to non-adiabatic following \cite{Weidemuller1994}. Once the atoms are back in the bright states, they can again undergo Sisyphus cooling before being pumped again in a dark state with, in average, a lower velocity.

We now introduce the main physical quantities at play to evaluate the efficiency and the limits of gray molasses (with $\Gamma$ the linewidth of the $^7S_3 \rightarrow {^7}P_2$ transition, $\lambda$ the laser wavelength, $\Omega$ the associated Rabi frequency and $\delta$ the detuning). The scattering rate in bright states or equivalently the optical pumping rate is given by $\Gamma' \propto \Omega^2/4\delta^2 \times \Gamma$ (with $\delta>\Gamma$). The capture velocity is $v_c=  \Gamma' / k$: only atoms with $v\leq v_c$ will be cooled down. The molasses temperature is $k_B T \propto \hbar \Omega^2 / \delta$, provided the associated \emph{r.m.s.} velocity is larger than $v_c$. As can be deduced from these equations, the lowest-limit temperature of atoms cooled within gray molasses is set by the AC Stark shift in the bright state. On the other hand, the capture velocity is controlled by the spontaneous emission rate in the bright state. For atoms to be efficiently captured in gray molasses, it is especially important for their velocity to be below $v_c$, since the light is blue detuned, and the traditional Doppler mechanism thus generates heating. Below $v_c$ however, the friction force of the Sisyphus cooling overcomes this heating process. The ultimate limit of cooling is reached when the capture velocity $v_c$ approaches the velocity associated with the \emph{r.m.s.} temperature of the cloud $\hbar \Omega^2 / \delta$. This sets a minimum laser power (or maximum $\delta$). Keeping in mind that it is required for molasses to be efficient that $\delta > \Gamma$, the lowest achievable temperature in gray molasses is such that $k_B T^*$ is of the order of five to ten $E_R$ as has been verified in numerous previous experiments with alkali atoms (see attached Table \ref{SotA}).

\section{Experimental setup}

\begin{figure}[]
	\centering
	\includegraphics[width=0.9\columnwidth]{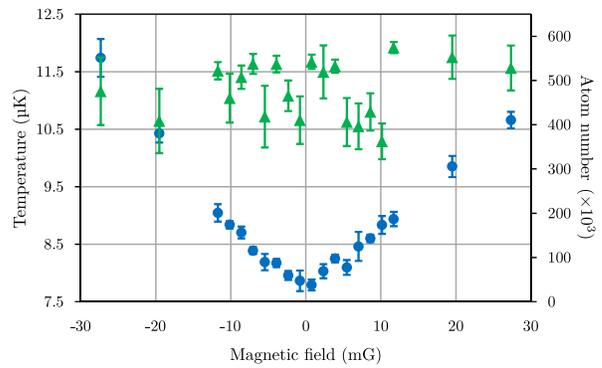}
	\caption{Evolution of the temperature (blue dots) and atom number (green triangles) with respect to the magnetic field in one direction of the horizontal plane in the absence of the ODT. The molasses phase duration is 3.5 ms, the intensity is ramped down after the initial 0.5 ms as described in the text and the detuning is set at $\delta=9.4\ \Gamma$. A similar behavior is observed in the other two directions of space.}
	\label{graphes3}
\end{figure}

Our experimental setup has been described in \cite{Bismut2011}. We first run a standard magneto-optical trap, with $\lambda=425$ nm laser beams, red detuned compared to the $^7S_3 \rightarrow {^7}P_4$ electronic transition. We superimpose on this trap a one-beam optical dipole trap provided by a far red-detuned $\lambda_{ODT}=1075$ nm laser beam propagating along a horizontal direction. To optimize loading of the trap, we apply a depumping laser beam at $427$ nm, close to the $^7S_3 \rightarrow {^7}P_3$ electronic transition. The combination of this light and the MOT light optically pumps atoms towards metastable dark states $^5S_2$, $^5D_4$, $^5D_3$ and $^5D_2$. These states are insensitive to the laser-cooling light, but they are sensitive to the AC Stark shift associated with the 1075 nm light. Optimum loading is realized when the depumping rate, mostly controlled by the 427 nm laser intensity, is fast compared to the light-assisted collision rate, and slow compared to the MOT equilibration time. After typically 200 ms of accumulation in the metastable states in presence of radio-frequency sweeps such that atoms in metastable states do not experience a magnetic field gradient \cite{Beaufils2008RF}, we turn the MOT beams and MOT magnetic-field gradient off, and repump all atoms from $^5S_2$, $^5D_4$ and $^5D_3$ into the electronic ground state using three laser diodes running at $633$, $654$ and $663$ nm.

This sequence ends up with typically $4.10^5$ atoms in the electronic ground state $^7S_3$, loaded in the optical trap at a temperature of $\approx60$ $\mu$K. This temperature, experimentally found to be $\approx 1/3$ of the trap depth \cite{Chicireanu2007}, arises from a compromise between the Doppler temperature and evaporation/spilling of the hottest atoms from the ODT. This sample is the starting point for our gray molasses experiment. Note that for this work we used smaller atom numbers than in our typical experiments, simply to increase the oven lifetime.

Gray molasses are produced using light derived from a Ti:Sa laser running at $858$ nm, and frequency doubled in a resonant bow-tie Fabry-P\'erot cavity including a LBO crystal. After injection of an optical fiber, up to $250$ mWatt are available for laser cooling. The beam is then split into one retro-reflected vertical beam and one retro-reflected horizontal beam in a butterfly shape providing the four horizontal molasses beams (same configuration as the MOT beams described in \cite{Chicireanu2006}). We then have a maximum total intensity of $2150\pm150$ mW.cm$^{-2}$ on the atoms which corresponds to a mean intensity per beam $I_{max}=358\pm25$ mWatt.cm$^{-2}=43\pm3$ $I_{sat}$ (where $I_{sat}=8.32$ mWatt.cm$^{-2}$ is the saturation intensity of the molasses transition). The uncertainty on the intensity is dominated by the uncertainty on the laser waists at the atoms location. The laser frequency is stabilized using an ultra-stable Fabry-P\'erot cavity, and the absolute frequency of the laser, determined by monitoring the fluorescence of the MOT in presence of the $429$ nm light, is controlled close to the $^7S_3 \rightarrow {^7}P_2$ transition using acousto-optic modulators.

\section{gray molasses optimization}

\begin{figure}[]
	\centering
	\includegraphics[width=0.9\columnwidth]{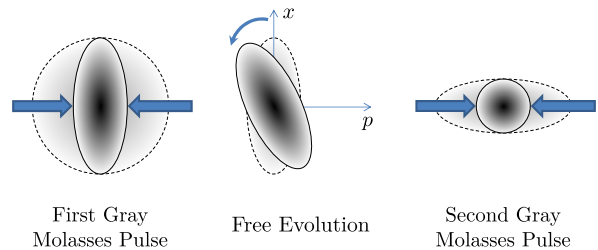}
	\caption{Scheme to cool all degrees of freedom in the ODT. The first molasses pulse reduces the width of the momentum distribution of the atoms. These are then left to evolve in the ODT for a quarter of the trap period, leading to a $\pi/2$ rotation in phase space. Then the second molasses pulse is performed. In each panel, the initial phase-space distribution ($p$,$x$) is represented by a shaded area with a dashed outline and the final one by an area with a solid outline. The arrows represent the effect of the GM pulses on the momentum distribution.}
	\label{principe}
\end{figure}

First we vary the detuning $\delta$ of the gray molasses laser compared to the $^7S_3 \rightarrow {^7}P_2$ transition. The intensity is kept at the maximum value of $43\ I_{sat}$ and the molasses duration is $0.5$ ms. We do not see any decrease in the final temperature for longer times. We perform this experiment both within the ODT, and just after the ODT is turned off (Fig.~\ref{graphes1}). The temperature is obtained by measuring the size of the atomic cloud in the vertical direction following a time-of-flight expansion after being released from the molasses (and the ODT when present). We find very similar efficiencies of the gray molasses in either case. The optimum detuning is slightly shifted from $\delta=2\pi\times47.3$ MHz $=9.4 \Gamma$ (with $\Gamma=2\pi\times5.029$ MHz) to $\delta=2\pi\times42.7$ MHz $=8.5 \Gamma$ when the trap is turned on, which is attributed to the fact that the ground and excited states do not undergo the same AC Stark shift at 1075 nm. Such a differential light shift is confirmed by numerical estimates \cite{Chicireanu2007}, and by the observation that, during the MOT stage, atoms in the ODT region undergo stronger fluorescence than away from the ODT, despite the fact that only a negligible fraction of $^7S_3$ atoms are trapped in the ODT.

Because the capture velocity $v_c$ is small, it is necessary to use large laser intensities to capture all atoms precooled in the MOT. However, once the atoms are captured and cooled, since the molasses temperature is set by $\propto  \Omega^2 / \delta$, it is favorable to ramp down $\Omega$ or ramp up $\delta$ in order to further reduce the temperature of the atoms until the ultimate limit of cooling $T^*$ is reached. Ramping down the intensity during the 0.5 ms molasses pulse does not lead to a decrease of the molasses temperature. For longer molasses durations however, we found it was beneficial to ramp down the intensity after the initial $0.5$ ms. For a total duration of $3.5$ ms and a ramp down of a few $I_{sat}$, the temperature decreased from $12.2$ to $11.5$ $\mu$K.

For efficient Sisyphus cooling, it is also important to lower the magnetic field amplitudes in all three directions of space. Since atoms are trapped in an optical dipole trap in our experiment, the dynamical control of the fields can be performed without losing atoms and without the reduction in density which inevitably happens when the magnetic field gradient of the MOT is turned off while the atomic cloud expands in free fall.
Away from the null magnetic field, we find that the temperature after the gray molasses pulse increases with $B$, with a sensitivity of $\approx 0.14$ $\mu$K.mGauss$^{-1}$. When the magnetic field components are compensated in all three directions to less than $\approx 3$ mGauss, we find an optimum temperature of $7.7\pm0.1$ $\mu$K (see Fig.~\ref{graphes3}) with a phase-space density of $(1.4\pm0.7)\times 10^{-4}$.

\section{Cooling all degrees of freedom in a trap}

\begin{figure}[]
	\centering
	\includegraphics[width=0.9\columnwidth]{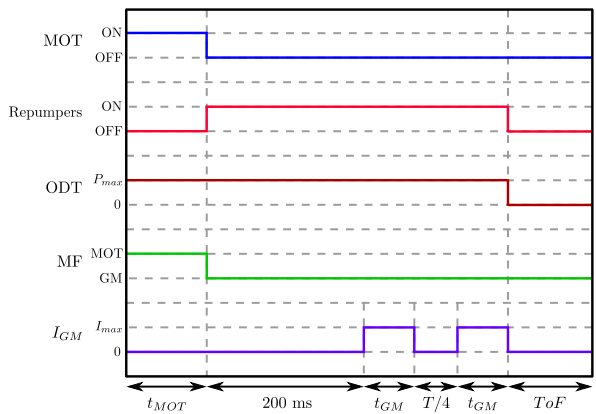}
	\caption{Time sequence for the simplest implementation of gray molasses and spatial compression, with two gray molasses pulses separated by a quarter-period of the optical dipolar trap in one direction. The 200 ms are necessary to repump the atoms from the metastable states. They also allow for the magnetic field (MF) to stabilize to its GM value.  $I_{GM}$ represents the GM laser beams mean intensity.}
	\label{exp_seq}
\end{figure}

\begin{figure}[]
	\centering
	\includegraphics[width=0.9\columnwidth]{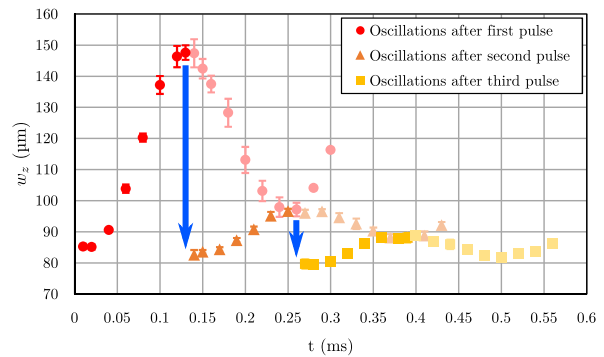}
	\caption{Evolution of the atomic cloud size (half-width at 1/e) in the vertical $z$ direction (after a time-of-flight) following a molasses pulse of 0.5 ms (red dots) or two (resp. three) molasses pulses separated by a quarter of the trap period $\frac{T_z}{4}\approx 0.13$ ms (orange triangles resp. yellow squares) and a wait time of duration $t$ in the ODT. The second and third molasses pulses are indicated by the downward arrows.}
	\label{graphes4}
\end{figure}

\begin{figure}[]
	\centering
	\includegraphics[width=0.9\columnwidth]{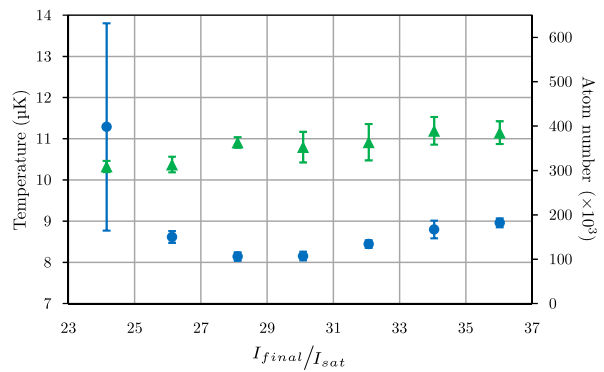}
	\caption{Temperature (blue dots) and atom number (green triangles) obtained when ramping down the molasses intensity from $36$ $I_{sat}$ to $I_{final}$. (The first two pulses were also ramped down from 43 to 41 $\ I_{sat}$ and from 41 to 38 $I_{sat}$ in 0.5 ms each.)}
	\label{graphes5}
\end{figure}

Gray molasses cooling results in almost freezing the motion of the atoms, but provides no spatial compression. Typically, experiments therefore first perform gray molasses, before loading the atoms into a conservative trap which is mode-matched to the profile of the atomic cloud.
We follow an alternative approach inspired by \cite{DePue1999} combining gray molasses and spatial compression. We apply the gray molasses cooling to optically trapped atomic clouds, benefiting from the fact that gray molasses for Cr are as efficient for trapped atoms as for untrapped atoms. The principle is described in Fig.~\ref{principe} and the experimental sequence is given in Fig.~\ref{exp_seq}. A first 0.5 ms gray molasses pulse is used to strongly lower the kinetic energy of the atoms within the trap. This results in an out-of-equilibrium situation where the energy is not equally distributed between the kinetic and potential contributions. We then use the property that free evolution during a quarter of the trap period $T/4$ within a harmonic trap leads to an exchange between position and momentum. At $T/4$, atoms initially distributed in the cloud volume are concentrated at the bottom of the trap and their initial potential energy has been converted into kinetic energy. The final spatial dispersion is set by the initial velocity dispersion of the cloud, whereas the velocity dispersion is raised back to the value it had before the gray molasses pulse. A second gray molasses pulse is then used to lower again the kinetic energy with negligible modification of the atoms spatial distribution. This results in a cloud which is both cooled and compressed, i.e. a cooling in both position and momentum. We do find that the first gray molasses pulse is followed by a (breathing) oscillation of the radius of the atomic cloud in the vertical $z$ direction (after release from the ODT, where the atoms were kept for a duration $t$ after the molasses pulse, and time-of-flight) which is greatly reduced after a second molasses pulse at $T/4$. The breathing motion is still not perfectly frozen however, presumably due to the anharmonicity of the trap. We thus apply a third pulse and indeed find that the amplitude of the oscillation is reduced further (Fig.~\ref{graphes4}). At the end of this sequence, the velocity and spatial distributions are cooled along the $z$ direction of space.

\begin{figure}[]
	\centering
	\includegraphics[width=0.9\columnwidth]{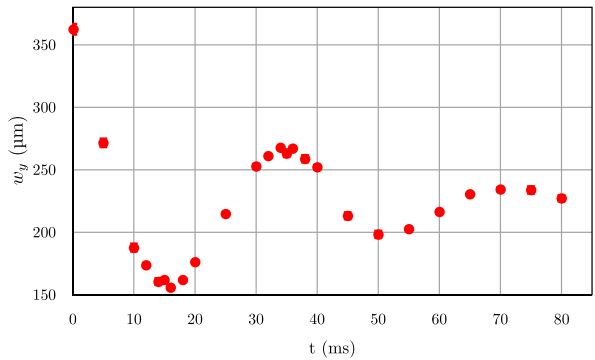}
	\caption{Evolution of the atomic cloud size in the $y$ horizontal directions (after a time-of-flight) following three molasses pulses of 0.5 ms separated by 0.13 ms (a quarter of the trap period in the vertical direction) and a wait time of duration $t$ in the ODT. We obtain a trap period in this direction of $\frac{T_y}{4}\approx16$ ms. At this point, the size of the cloud is minimum and the velocity dispersion is maximum. Another GM pulse is then applied.}
	\label{graphes6}
\end{figure}

\begin{figure}[]
	\centering
	\includegraphics[width=0.9\columnwidth]{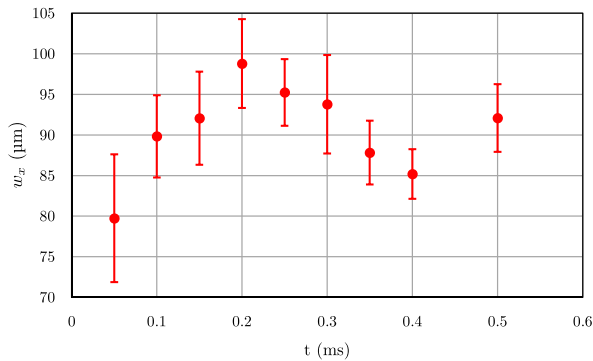}
	\caption{Evolution of the atomic cloud size in the $x$ horizontal direction (after a time-of-flight) following four molasses pulses of 0.5 ms separated by 16 ms (a quarter of the trap period in the $y$ horizontal direction) for the first two and by 0.13 ms (a quarter of the trap period in the $z$ direction) for the last three and a wait time of duration $t$ in the ODT. The trap period in this direction is $\frac{T_x}{4}\approx0.2$ ms.}
	\label{graphes7}
\end{figure}

To reduce the temperature further we then ramp down the intensity during the third pulse. This time we keep the slope of the ramp constant to $-3.9$ $I_{sat}$ ms${-1}$ and vary the duration of the pulse. An optimum is obtained for a 2.3 ms pulse and a final intensity of $29\ I_{sat}$ leading to a final temperature of $8.1\pm0.1$ $\mu$K (see Fig.~\ref{graphes5}).

We then perform two more molasses pulses to have the same cooling and compression in the two horizontal directions of space $y$ (Fig.~\ref{graphes6}) and $x$ (Fig.~\ref{graphes7}). We find that the very elongated direction strongly decouples from the two other directions. In practice, when the two directions with the highest and lowest trapping frequencies ($z$ and $y$) are cooled in space and momentum, so is the third one ($x$).

Taken together, four pulses are then sufficient to cool and compress the whole sample. A reduction of the temperature from 60 to 8 $\mu$K affecting only the kinetic energy would result in an increase of a factor $\eta_1=20$ in phase-space density. Our improved scheme where gray molasses pulses are interspersed with free evolution in the trap instead leads to a much larger gain in phase-space density, of $\eta_2=250$ (see Fig.~\ref{PSD}). In a perfect trap, one expects to reach $\eta_2=\eta_1^2$. The discrepancy between $\eta_2$  and $\eta_1^2$  is attributed to anharmonicity.

In practice, after a sequence of total duration 20 ms, we end up with a thermal sample consisting of $\approx 4.10^5$ atoms at a temperature of $\approx8.1$ $\mu$K at a high phase-space density of $\approx(1.7\pm0.9)\times10^{-3}$ in an optical-dipole trap of frequencies about 1.3 kHz and 1.9 kHz along the strongly confining axes and 14 Hz along the weakly confining axis.

The characteristic collision times at the beginning and at the end of the cooling sequence are $(n\sigma v)_{initial}^{-1}\approx94$ ms and $(n\sigma v)_{final}^{-1}\approx21$ ms, we therefore think that collisions can be neglected for the 20 ms of the cooling sequence. We can compare our results to those obtained in  \cite{Volochkov2014} with demagnetization cooling of chromium in an optical dipole trap, where a temperature of 6 $\mu$K and a phase space density of $\approx0.02$ were reached, albeit at the cost of atom losses due to much longer cooling times of the order of 10 s.

\section{Conclusion}

We have combined gray molasses cooling and rotation in phase-space to produce ultracold chromium atomic thermal ensembles of high phase-space density. Such ensembles stand as very favorable starting points for further cooling down to the quantum regime using for example forced evaporation in a crossed beam optical trap. Furthermore the presented scheme could be of benefit for other atomic species when gray molasses are possible within an ODT.

\vspace{1cm}

\begin{acknowledgments}
	
We acknowledge financial support from Conseil R\'egional d'Ile-de-France under DIM Nano-K/IFRAF, CNRS, Minist\`ere de l'Enseignement Sup\'erieur et de la Recherche within CPER Contract, Universit\'e Sorbonne Paris Cit\'e, and the Indo-French Centre for the Promotion of Advanced Research under LORIC5404-1 and PPKC contracts.

\end{acknowledgments}

\onecolumngrid
\renewcommand{\arraystretch}{1.3}
\setlength{\tabcolsep}{1.0mm}

\begin{sidewaystable}[]
	\centering
	\begin{threeparttable}
	\begin{tabular}{c|c|c|c|c|c|c|c|c|c|c|c}
		Element                                                                                        & Paper                                                                                     & Transition                                                                                     & \begin{tabular}[c]{@{}c@{}}Laser\\[-1.5mm] configuration\end{tabular}                                        & \begin{tabular}[c]{@{}c@{}}\large ${}^{\Gamma}{\mskip -3mu/\mskip -0mu}_{2\pi}$\\[-1.5mm] \scriptsize (MHz)\end{tabular}    & \begin{tabular}[c]{@{}c@{}}$T_{rec}$\\[-1.5mm] \scriptsize ($\mu$K)\end{tabular}    & \begin{tabular}[c]{@{}c@{}}$I_{sat}$\\[-1.5mm] \scriptsize (mW/cm$^2$)\end{tabular}    & \begin{tabular}[c]{@{}c@{}}Duration\\[-1.5mm] \scriptsize (ms)\end{tabular}     & \large ${}^{\delta}{\mskip -3mu/\mskip -0mu}_{\Gamma}$                                                                                 & \large ${}^{I}{\mskip -3mu/\mskip -0mu}_{I_{sat}}$                                                                                                                                & \large ${}^{T}{\mskip -3mu/\mskip -0mu}_{T_{rec}}$     & $P.S.D.$                                                                                                                             \\ \hline
		\begin{tabular}[c]{@{}c@{}}$^{133}Cs$\\[-1.5mm] (Boson)\end{tabular}                             & \begin{tabular}[c]{@{}c@{}}1999\\[-1.5mm] Trich\'e\end{tabular} \cite{Triche1999}           & \scriptsize $D_2\ {^2S_{1/2}},F=3 \rightarrow {^2P_{3/2}},F'=2$                                & \begin{tabular}[c]{@{}c@{}}Four-beam\\[-1.5mm]  $lin\bot lin$\end{tabular}                                   & $5.234$                                                                                                                   & $0.198$                                                                                       & $1.1$                                                                                & $\sim10$                                                                                      & $4$                                                                                                                                    & \large $^{\sim1}\hspace{-1mm}\searrow_{\, 0.1}$                                                                                                                                   & $4$                                                    & \o                                                                                                                                   \\ \hline
		\multirow{2}{*}[-2mm]{\begin{tabular}[c]{@{}c@{}}$^{39}K$\\[-1.5mm] (Boson)\end{tabular}}        & \begin{tabular}[c]{@{}c@{}}2013\\[-1.5mm] Nath\end{tabular} \cite{Nath2013}                 & \multirow{2}{*}[-2mm]{\scriptsize $D_1\ {^2S_{1/2}},F=2 \rightarrow {^2P_{1/2}},F'=2$}         & \multirow{2}{*}[-2mm]{\begin{tabular}[c]{@{}c@{}}Six-beam\\[-1.5mm]  $\sigma^+-\ \sigma^-$\end{tabular}}     & \multirow{2}{*}[-2mm]{$5.956$}                                                                                            & \multirow{2}{*}[-2mm]{$0.829$}                                                                & \multirow{2}{*}[-2mm]{$1.7$}                                                         & $6$                                                                                           & $5$                                                                                                                                    & $3.5$                                                                                                                                                                             & $14.5$                                                 & $>2\times 10^{-5}$                                                                                                                   \\
		& \begin{tabular}[c]{@{}c@{}}2013\\[-1.5mm] Salomon\end{tabular} \cite{Salomon2013}           &                                                                                                &                                                                                                            &                                                                                                                           &                                                                                               &                                                                                      & $7$                                                                                           & $3.5$                                                                                                                                  & \large $^{3.5}\hspace{-1mm}\searrow _{\, 0.2}$                                                                                                                                    & $7.2$                                                  & $2\times 10^{-4}$                                                                                                                    \\ \hline
		\multirow{2}{*}[-2mm]{\begin{tabular}[c]{@{}c@{}}$^{40}K$\\[-1.5mm] (Fermion)\end{tabular}}      & \begin{tabular}[c]{@{}c@{}}2015\\[-1.5mm] Sievers\end{tabular} \cite{Sievers2015}           & \multirow{2}{*}[-2mm]{\scriptsize $D_1\ {^2S_{1/2}},F=9/2 \rightarrow {^2P_{1/2}},F'=7/2$}     & \multirow{2}{*}[-2mm]{\begin{tabular}[c]{@{}c@{}}Six-beam\\[-1.5mm] $\sigma^+-\ \sigma^-$\end{tabular}}      & \multirow{2}{*}[-2mm]{$5.956$}                                                                                            & \multirow{2}{*}[-2mm]{$0.808$}                                                                & \multirow{2}{*}[-2mm]{$1.7$}                                                         & $5$                                                                                           & $2.3$                                                                                                                                  & $14$                                                                                                                                                                              & $13.6$                                                 & $1.7\times 10^{-4}$                                                                                                                  \\
		& \begin{tabular}[c]{@{}c@{}}2015\\[-1.5mm] Tarnowski\end{tabular} \cite{Tarnowski2015}       &                                                                                                &                                                                                                            &                                                                                                                           &                                                                                               &                                                                                      & $13$                                                                                          & $4.5$                                                                                                                                  & \large $^{17}\hspace{-1mm}\searrow _{\, 3.4}$                                                                                                                                     & $7.0$                                                  & $1.5\times 10^{-5}$                                                                                                                  \\ \hline
		\multirow{2}{*}[-2mm]{\begin{tabular}[c]{@{}c@{}}$^{41}K$\\[-1.5mm] (Boson)\end{tabular}}        & \begin{tabular}[c]{@{}c@{}}2015\\[-1.5mm] Sievers\end{tabular} \cite{Sievers2015}           & \scriptsize $D_1$                                                                              & \o                                                                                                         & \multirow{2}{*}[-2mm]{$5.956$}                                                                                            & \multirow{2}{*}[-2mm]{$0.788$}                                                                & \multirow{2}{*}[-2mm]{$1.7$}                                                         & \o                                                                                            & \o                                                                                                                                     & \o                                                                                                                                                                                & $25.4$                                                 & $1.1\times 10^{-4}$                                                                                                                  \\
		& \begin{tabular}[c]{@{}c@{}}2016\\[-1.5mm] Chen\end{tabular} \cite{Chen2016}                 & \scriptsize $D_1\ {^2S_{1/2}},F=2 \rightarrow {^2P_{1/2}},F'=2$                                & \begin{tabular}[c]{@{}c@{}}Six-beam\\[-1.5mm] $\sigma^+-\ \sigma^-$\end{tabular}                             &                                                                                                                           &                                                                                               &                                                                                      & $17$                                                                                          & $5.3$                                                                                                                                  & \large $^{12.5}\hspace{-1mm}\searrow _{\, 1.5}$                                                                                                                                   & $53.3$                                                 & $5.4\times 10^{-6}$                                                                                                                  \\ \hline
		\multirow{2}{*}[-2mm]{\begin{tabular}[c]{@{}c@{}}$^{6}Li$\\[-1.5mm] (Fermion)\end{tabular}}      & \begin{tabular}[c]{@{}c@{}}2014\\[-1.5mm] Burchianti\end{tabular} \cite{Burchianti2014}     & \multirow{2}{*}[-2mm]{\scriptsize $D_1\ {^2S_{1/2}},F=3/2 \rightarrow {^2P_{1/2}},F'=3/2$}     & \multirow{2}{*}[-2mm]{\begin{tabular}[c]{@{}c@{}}Six-beam\\[-1.5mm] $\sigma^+-\ \sigma^-$\end{tabular}}      & \multirow{2}{*}[-2mm]{$5.872$}                                                                                            & \multirow{2}{*}[-2mm]{$7.071$}                                                                & \multirow{2}{*}[-2mm]{$2.54$}                                                        & $2$                                                                                           & $5.4$                                                                                                                                  & $2.7$                                                                                                                                                                             & $5.7$                                                  & $2\times 10^{-5}$                                                                                                                    \\
		& \begin{tabular}[c]{@{}c@{}}2015\\[-1.5mm] Sievers\end{tabular} \cite{Sievers2015}           &                                                                                                &                                                                                                            &                                                                                                                           &                                                                                               &                                                                                      & $5$                                                                                           & $4$                                                                                                                                    & \large $^{14.6}\hspace{-1mm}\searrow _{\, 2.5}$                                                                                                                                   & $6.2$                                                  & $\sim10^{-4}$                                                                                                                        \\ \hline
		\begin{tabular}[c]{@{}c@{}}$^{7}Li$\\[-1.5mm] (Boson)\end{tabular}                               & \begin{tabular}[c]{@{}c@{}}2013\\[-1.5mm] Grier\end{tabular} \cite{Grier2013}               & \scriptsize $D_1\ {^2S_{1/2}},F=2 \rightarrow {^2P_{1/2}},F'=2$                                & \begin{tabular}[c]{@{}c@{}}Six-beam\\[-1.5mm] $\sigma^+-\ \sigma^-$\end{tabular}                             & $5.872$                                                                                                                   & $6.063$                                                                                       & $2.54$                                                                               & $2$                                                                                           & $4.5$                                                                                                                                  & $\gtrsim45$                                                                                                                                                                       & $9.9$                                                  & $\sim10^{-5}$                                                                                                                        \\ \hline
		\begin{tabular}[c]{@{}c@{}}$^{23}Na$\\[-1.5mm] (Boson)\end{tabular}                              & \begin{tabular}[c]{@{}c@{}}2016\\[-1.5mm] Colzi\end{tabular} \cite{Colzi2016}               & \scriptsize $D_1\ {^2S_{1/2}},F=2 \rightarrow {^2P_{1/2}},F'=2$                                & \begin{tabular}[c]{@{}c@{}}Six-beam\\[-1.5mm] $\sigma^+-\ \sigma^-$\end{tabular}                             & $9.765$                                                                                                                   & $2.395$                                                                                       & $6.26$                                                                               & $15.5$                                                                                        & \begin{tabular}[c]{@{}c@{}}$5$ \scriptsize (0.5 ms)\\[-1.5mm] $12$ \scriptsize (15 ms)\end{tabular}    & \begin{tabular}[c]{@{}c@{}}$13.3$ \scriptsize (0.5 ms)\\[-1.5mm] \large $^{8.3}\hspace{-1mm}\searrow _{\, 6.7}$ \scriptsize (15 ms)\end{tabular}      & $3.7$                                                  & $1.1\times 10^{-4}$                                                                                                                  \\ \hline
		\begin{tabular}[c]{@{}c@{}}$^{87}Rb$\\[-1.5mm] (Boson)\end{tabular}                              & \begin{tabular}[c]{@{}c@{}}2018\\[-1.5mm] Rosi\end{tabular} \cite{Rosi2018}                 & \scriptsize $D_2\ {^2S_{1/2}},F=2 \rightarrow {^2P_{3/2}},F'=2$                                & \begin{tabular}[c]{@{}c@{}}Six-beam\\[-1.5mm] $\sigma^+-\ \sigma^-$\end{tabular}                             & $6.065$                                                                                                                   & $0.362$                                                                                       & $1.67$                                                                               & $3$                                                                                           & $12$                                                                                                                                   & $\sim6$ (ramped)                                                                                                                                                                  & $11.0$                                                 & $4.0\times 10^{-6}$                                                                                                                  \\ \hline
		\begin{tabular}[c]{@{}c@{}}$^{4}He^*$\\[-1.5mm] (Boson)\end{tabular}                             & \begin{tabular}[c]{@{}c@{}}2015\\[-1.5mm] Bouton\end{tabular} \cite{Bouton2015}             & \scriptsize ${2\ {^3S_{1}}} \rightarrow {2\ {^3P_{1}}}$                                        & \o                                                                                                         & $1.626$                                                                                                                   & $4.077$                                                                                       & $0.16$                                                                               & $5$                                                                                           & $10$                                                                                                                                   & $20$                                                                                                                                                                              & $4.9$                                                  & $7.8\times 10^{-5}$                                                                                                                  \\ \hline		
		\begin{tabular}[c]{@{}c@{}}$^{52}Cr$\\[-1.5mm] (Boson)\end{tabular}                              & This paper                                                                                & \scriptsize ${^7S_{3}} \rightarrow {^7P_{2}}$                                                  & \begin{tabular}[c]{@{}c@{}}Six-beam\\[-1.5mm] $\sigma^+-\ \sigma^-$\end{tabular}                             & $5.029$                                                                                                                   & $2.002$                                                                                       & $8.32$                                                                               & $3.5$                                                                                         & $3.8$                                                                                                                                  & \large $^{43}\hspace{-1mm}\searrow _{\, \sim35}$                                                                                                                                  & $3.9$                                                  & \begin{tabular}[c]{@{}c@{}}$1.4\times10^{-4}$\\[-1.5mm] or $1.7\times10^{-3}$~\tnote{*}\end{tabular}
	\end{tabular}
	\begin{tablenotes}
	\item[*] When using the compression scheme.
	\end{tablenotes}
	\end{threeparttable}
	\caption{State of the art for grey molasses cooling of alkali atoms, metastable helium and chromium. The results presented here are those achieving the lowest temperatures. Linewidths were taken from NIST atomic spectra database \cite{NIST_ASD} or from the reference data compiled on Daniel Steck's website \cite{Alkali_data}. Recoil temperatures are obtained using the relation $T_{rec}=\frac{h^2}{m k_B \lambda^2}$. Saturating intensities are obtained with the relation $I_{sat}=\frac{\pi h c \Gamma}{30 \lambda^3}$. The phase space densities were either taken from the cited papers or calculated using the relation $P.S.D. = n \lambda^3_{dB}$ with the thermal de Broglie wavelength $\lambda_{dB} = h/\sqrt{2 \pi m k_B T}$. The parameters given for our paper are for a single molasses pulse without the compression part. The final $P.S.D.$ for four gray molasses pulses interspaced by free evolution within the trap is also given.}
	\label{SotA}
\end{sidewaystable}

\twocolumngrid
\clearpage

\bibliography{GM_biblio}

\begin{thebibliography}{44}%
\makeatletter
\providecommand \@ifxundefined [1]{%
 \@ifx{#1\undefined}
}%
\providecommand \@ifnum [1]{%
 \ifnum #1\expandafter \@firstoftwo
 \else \expandafter \@secondoftwo
 \fi
}%
\providecommand \@ifx [1]{%
 \ifx #1\expandafter \@firstoftwo
 \else \expandafter \@secondoftwo
 \fi
}%
\providecommand \natexlab [1]{#1}%
\providecommand \enquote  [1]{``#1''}%
\providecommand \bibnamefont  [1]{#1}%
\providecommand \bibfnamefont [1]{#1}%
\providecommand \citenamefont [1]{#1}%
\providecommand \href@noop [0]{\@secondoftwo}%
\providecommand \href [0]{\begingroup \@sanitize@url \@href}%
\providecommand \@href[1]{\@@startlink{#1}\@@href}%
\providecommand \@@href[1]{\endgroup#1\@@endlink}%
\providecommand \@sanitize@url [0]{\catcode `\\12\catcode `\$12\catcode
  `\&12\catcode `\#12\catcode `\^12\catcode `\_12\catcode `\%12\relax}%
\providecommand \@@startlink[1]{}%
\providecommand \@@endlink[0]{}%
\providecommand \url  [0]{\begingroup\@sanitize@url \@url }%
\providecommand \@url [1]{\endgroup\@href {#1}{\urlprefix }}%
\providecommand \urlprefix  [0]{URL }%
\providecommand \Eprint [0]{\href }%
\providecommand \doibase [0]{http://dx.doi.org/}%
\providecommand \selectlanguage [0]{\@gobble}%
\providecommand \bibinfo  [0]{\@secondoftwo}%
\providecommand \bibfield  [0]{\@secondoftwo}%
\providecommand \translation [1]{[#1]}%
\providecommand \BibitemOpen [0]{}%
\providecommand \bibitemStop [0]{}%
\providecommand \bibitemNoStop [0]{.\EOS\space}%
\providecommand \EOS [0]{\spacefactor3000\relax}%
\providecommand \BibitemShut  [1]{\csname bibitem#1\endcsname}%
\let\auto@bib@innerbib\@empty
\bibitem [{\citenamefont {Griesmaier}\ \emph {et~al.}(2005)\citenamefont
  {Griesmaier}, \citenamefont {Werner}, \citenamefont {Hensler}, \citenamefont
  {Stuhler},\ and\ \citenamefont {Pfau}}]{Griesmaier2005}%
  \BibitemOpen
  \bibfield  {author} {\bibinfo {author} {\bibfnamefont {A.}~\bibnamefont
  {Griesmaier}}, \bibinfo {author} {\bibfnamefont {J.}~\bibnamefont {Werner}},
  \bibinfo {author} {\bibfnamefont {S.}~\bibnamefont {Hensler}}, \bibinfo
  {author} {\bibfnamefont {J.}~\bibnamefont {Stuhler}}, \ and\ \bibinfo
  {author} {\bibfnamefont {T.}~\bibnamefont {Pfau}},\ }\href {\doibase
  10.1103/PhysRevLett.94.160401} {\bibfield  {journal} {\bibinfo  {journal}
  {Phys. Rev. Lett.}\ }\textbf {\bibinfo {volume} {94}},\ \bibinfo {pages}
  {160401} (\bibinfo {year} {2005})}\BibitemShut {NoStop}%
\bibitem [{\citenamefont {Beaufils}\ \emph
  {et~al.}(2008{\natexlab{a}})\citenamefont {Beaufils}, \citenamefont
  {Chicireanu}, \citenamefont {Zanon}, \citenamefont {Laburthe-Tolra},
  \citenamefont {Mar\'echal}, \citenamefont {Vernac}, \citenamefont {Keller},\
  and\ \citenamefont {Gorceix}}]{Beaufils2008}%
  \BibitemOpen
  \bibfield  {author} {\bibinfo {author} {\bibfnamefont {Q.}~\bibnamefont
  {Beaufils}}, \bibinfo {author} {\bibfnamefont {R.}~\bibnamefont
  {Chicireanu}}, \bibinfo {author} {\bibfnamefont {T.}~\bibnamefont {Zanon}},
  \bibinfo {author} {\bibfnamefont {B.}~\bibnamefont {Laburthe-Tolra}},
  \bibinfo {author} {\bibfnamefont {E.}~\bibnamefont {Mar\'echal}}, \bibinfo
  {author} {\bibfnamefont {L.}~\bibnamefont {Vernac}}, \bibinfo {author}
  {\bibfnamefont {J.-C.}\ \bibnamefont {Keller}}, \ and\ \bibinfo {author}
  {\bibfnamefont {O.}~\bibnamefont {Gorceix}},\ }\href {\doibase
  10.1103/PhysRevA.77.061601} {\bibfield  {journal} {\bibinfo  {journal} {Phys.
  Rev. A}\ }\textbf {\bibinfo {volume} {77}},\ \bibinfo {pages} {061601}
  (\bibinfo {year} {2008}{\natexlab{a}})}\BibitemShut {NoStop}%
\bibitem [{\citenamefont {Lu}\ \emph {et~al.}(2011)\citenamefont {Lu},
  \citenamefont {Burdick}, \citenamefont {Youn},\ and\ \citenamefont
  {Lev}}]{Lu2011}%
  \BibitemOpen
  \bibfield  {author} {\bibinfo {author} {\bibfnamefont {M.}~\bibnamefont
  {Lu}}, \bibinfo {author} {\bibfnamefont {N.~Q.}\ \bibnamefont {Burdick}},
  \bibinfo {author} {\bibfnamefont {S.~H.}\ \bibnamefont {Youn}}, \ and\
  \bibinfo {author} {\bibfnamefont {B.~L.}\ \bibnamefont {Lev}},\ }\href
  {\doibase 10.1103/PhysRevLett.107.190401} {\bibfield  {journal} {\bibinfo
  {journal} {Phys. Rev. Lett.}\ }\textbf {\bibinfo {volume} {107}},\ \bibinfo
  {pages} {190401} (\bibinfo {year} {2011})}\BibitemShut {NoStop}%
\bibitem [{\citenamefont {Aikawa}\ \emph {et~al.}(2012)\citenamefont {Aikawa},
  \citenamefont {Frisch}, \citenamefont {Mark}, \citenamefont {Baier},
  \citenamefont {Rietzler}, \citenamefont {Grimm},\ and\ \citenamefont
  {Ferlaino}}]{Aikawa2012}%
  \BibitemOpen
  \bibfield  {author} {\bibinfo {author} {\bibfnamefont {K.}~\bibnamefont
  {Aikawa}}, \bibinfo {author} {\bibfnamefont {A.}~\bibnamefont {Frisch}},
  \bibinfo {author} {\bibfnamefont {M.}~\bibnamefont {Mark}}, \bibinfo {author}
  {\bibfnamefont {S.}~\bibnamefont {Baier}}, \bibinfo {author} {\bibfnamefont
  {A.}~\bibnamefont {Rietzler}}, \bibinfo {author} {\bibfnamefont
  {R.}~\bibnamefont {Grimm}}, \ and\ \bibinfo {author} {\bibfnamefont
  {F.}~\bibnamefont {Ferlaino}},\ }\href {\doibase
  10.1103/PhysRevLett.108.210401} {\bibfield  {journal} {\bibinfo  {journal}
  {Phys. Rev. Lett.}\ }\textbf {\bibinfo {volume} {108}},\ \bibinfo {pages}
  {210401} (\bibinfo {year} {2012})}\BibitemShut {NoStop}%
\bibitem [{\citenamefont {Lahaye}\ \emph {et~al.}(2009)\citenamefont {Lahaye},
  \citenamefont {Menotti}, \citenamefont {Santos}, \citenamefont {Lewenstein},\
  and\ \citenamefont {Pfau}}]{Lahaye2009}%
  \BibitemOpen
  \bibfield  {author} {\bibinfo {author} {\bibfnamefont {T.}~\bibnamefont
  {Lahaye}}, \bibinfo {author} {\bibfnamefont {C.}~\bibnamefont {Menotti}},
  \bibinfo {author} {\bibfnamefont {L.}~\bibnamefont {Santos}}, \bibinfo
  {author} {\bibfnamefont {M.}~\bibnamefont {Lewenstein}}, \ and\ \bibinfo
  {author} {\bibfnamefont {T.}~\bibnamefont {Pfau}},\ }\href {\doibase
  10.1088/0034-4885/72/12/126401} {\bibfield  {journal} {\bibinfo  {journal}
  {Rep. Prog. Phys.}\ }\textbf {\bibinfo {volume} {72}},\ \bibinfo {pages}
  {126401} (\bibinfo {year} {2009})}\BibitemShut {NoStop}%
\bibitem [{\citenamefont {Baranov}(2008)}]{Baranov2008}%
  \BibitemOpen
  \bibfield  {author} {\bibinfo {author} {\bibfnamefont {M.~A.}\ \bibnamefont
  {Baranov}},\ }\href {\doibase 10.1016/j.physrep.2008.04.007} {\bibfield
  {journal} {\bibinfo  {journal} {Phys. Rep.}\ }\textbf {\bibinfo {volume}
  {464}},\ \bibinfo {pages} {71} (\bibinfo {year} {2008})}\BibitemShut
  {NoStop}%
\bibitem [{\citenamefont {Lahaye}\ \emph {et~al.}(2008)\citenamefont {Lahaye},
  \citenamefont {Metz}, \citenamefont {Fr\"ohlich}, \citenamefont {Koch},
  \citenamefont {Meister}, \citenamefont {Griesmaier}, \citenamefont {Pfau},
  \citenamefont {Saito}, \citenamefont {Kawaguchi},\ and\ \citenamefont
  {Ueda}}]{Lahaye2008}%
  \BibitemOpen
  \bibfield  {author} {\bibinfo {author} {\bibfnamefont {T.}~\bibnamefont
  {Lahaye}}, \bibinfo {author} {\bibfnamefont {J.}~\bibnamefont {Metz}},
  \bibinfo {author} {\bibfnamefont {B.}~\bibnamefont {Fr\"ohlich}}, \bibinfo
  {author} {\bibfnamefont {T.}~\bibnamefont {Koch}}, \bibinfo {author}
  {\bibfnamefont {M.}~\bibnamefont {Meister}}, \bibinfo {author} {\bibfnamefont
  {A.}~\bibnamefont {Griesmaier}}, \bibinfo {author} {\bibfnamefont
  {T.}~\bibnamefont {Pfau}}, \bibinfo {author} {\bibfnamefont {H.}~\bibnamefont
  {Saito}}, \bibinfo {author} {\bibfnamefont {Y.}~\bibnamefont {Kawaguchi}}, \
  and\ \bibinfo {author} {\bibfnamefont {M.}~\bibnamefont {Ueda}},\ }\href
  {\doibase 10.1103/PhysRevLett.101.080401} {\bibfield  {journal} {\bibinfo
  {journal} {Phys. Rev. Lett.}\ }\textbf {\bibinfo {volume} {101}},\ \bibinfo
  {pages} {080401} (\bibinfo {year} {2008})}\BibitemShut {NoStop}%
\bibitem [{\citenamefont {de~Paz}\ \emph {et~al.}(2013)\citenamefont {de~Paz},
  \citenamefont {Sharma}, \citenamefont {Chotia}, \citenamefont {Mar\'echal},
  \citenamefont {Huckans}, \citenamefont {Pedri}, \citenamefont {Santos},
  \citenamefont {Gorceix}, \citenamefont {Vernac},\ and\ \citenamefont
  {Laburthe-Tolra}}]{DePaz2013}%
  \BibitemOpen
  \bibfield  {author} {\bibinfo {author} {\bibfnamefont {A.}~\bibnamefont
  {de~Paz}}, \bibinfo {author} {\bibfnamefont {A.}~\bibnamefont {Sharma}},
  \bibinfo {author} {\bibfnamefont {A.}~\bibnamefont {Chotia}}, \bibinfo
  {author} {\bibfnamefont {E.}~\bibnamefont {Mar\'echal}}, \bibinfo {author}
  {\bibfnamefont {J.~H.}\ \bibnamefont {Huckans}}, \bibinfo {author}
  {\bibfnamefont {P.}~\bibnamefont {Pedri}}, \bibinfo {author} {\bibfnamefont
  {L.}~\bibnamefont {Santos}}, \bibinfo {author} {\bibfnamefont
  {O.}~\bibnamefont {Gorceix}}, \bibinfo {author} {\bibfnamefont
  {L.}~\bibnamefont {Vernac}}, \ and\ \bibinfo {author} {\bibfnamefont
  {B.}~\bibnamefont {Laburthe-Tolra}},\ }\href {\doibase
  10.1103/PhysRevLett.111.185305} {\bibfield  {journal} {\bibinfo  {journal}
  {Phys. Rev. Lett.}\ }\textbf {\bibinfo {volume} {111}},\ \bibinfo {pages}
  {185305} (\bibinfo {year} {2013})}\BibitemShut {NoStop}%
\bibitem [{\citenamefont {Chomaz}\ \emph {et~al.}(2018)\citenamefont {Chomaz},
  \citenamefont {van Bijnen}, \citenamefont {Petter}, \citenamefont {Faraoni},
  \citenamefont {Baier}, \citenamefont {Becher}, \citenamefont {Mark},
  \citenamefont {W\"achtler}, \citenamefont {Santos},\ and\ \citenamefont
  {Ferlaino}}]{Chomaz2018}%
  \BibitemOpen
  \bibfield  {author} {\bibinfo {author} {\bibfnamefont {L.}~\bibnamefont
  {Chomaz}}, \bibinfo {author} {\bibfnamefont {R.~M.~W.}\ \bibnamefont {van
  Bijnen}}, \bibinfo {author} {\bibfnamefont {D.}~\bibnamefont {Petter}},
  \bibinfo {author} {\bibfnamefont {G.}~\bibnamefont {Faraoni}}, \bibinfo
  {author} {\bibfnamefont {S.}~\bibnamefont {Baier}}, \bibinfo {author}
  {\bibfnamefont {J.~H.}\ \bibnamefont {Becher}}, \bibinfo {author}
  {\bibfnamefont {M.~J.}\ \bibnamefont {Mark}}, \bibinfo {author}
  {\bibfnamefont {F.}~\bibnamefont {W\"achtler}}, \bibinfo {author}
  {\bibfnamefont {L.}~\bibnamefont {Santos}}, \ and\ \bibinfo {author}
  {\bibfnamefont {F.}~\bibnamefont {Ferlaino}},\ }\href {\doibase
  10.1038/s41567-018-0054-7} {\bibfield  {journal} {\bibinfo  {journal} {Nat.
  Phys.}\ }\textbf {\bibinfo {volume} {14}},\ \bibinfo {pages} {442} (\bibinfo
  {year} {2018})}\BibitemShut {NoStop}%
\bibitem [{\citenamefont {Schmitt}\ \emph {et~al.}(2016)\citenamefont
  {Schmitt}, \citenamefont {Wenzel}, \citenamefont {B\"ottcher}, \citenamefont
  {Ferrier-Barbut},\ and\ \citenamefont {Pfau}}]{Schmitt2016}%
  \BibitemOpen
  \bibfield  {author} {\bibinfo {author} {\bibfnamefont {M.}~\bibnamefont
  {Schmitt}}, \bibinfo {author} {\bibfnamefont {M.}~\bibnamefont {Wenzel}},
  \bibinfo {author} {\bibfnamefont {F.}~\bibnamefont {B\"ottcher}}, \bibinfo
  {author} {\bibfnamefont {I.}~\bibnamefont {Ferrier-Barbut}}, \ and\ \bibinfo
  {author} {\bibfnamefont {T.}~\bibnamefont {Pfau}},\ }\href {\doibase
  10.1038/nature20126} {\bibfield  {journal} {\bibinfo  {journal} {Nature}\
  }\textbf {\bibinfo {volume} {539}},\ \bibinfo {pages} {259} (\bibinfo {year}
  {2016})}\BibitemShut {NoStop}%
\bibitem [{\citenamefont {Pasquiou}\ \emph {et~al.}(2012)\citenamefont
  {Pasquiou}, \citenamefont {Mar\'echal}, \citenamefont {Vernac}, \citenamefont
  {Gorceix},\ and\ \citenamefont {Laburthe-Tolra}}]{Pasquiou2012}%
  \BibitemOpen
  \bibfield  {author} {\bibinfo {author} {\bibfnamefont {B.}~\bibnamefont
  {Pasquiou}}, \bibinfo {author} {\bibfnamefont {E.}~\bibnamefont
  {Mar\'echal}}, \bibinfo {author} {\bibfnamefont {L.}~\bibnamefont {Vernac}},
  \bibinfo {author} {\bibfnamefont {O.}~\bibnamefont {Gorceix}}, \ and\
  \bibinfo {author} {\bibfnamefont {B.}~\bibnamefont {Laburthe-Tolra}},\ }\href
  {\doibase 10.1103/PhysRevLett.108.045307} {\bibfield  {journal} {\bibinfo
  {journal} {Phys. Rev. Lett.}\ }\textbf {\bibinfo {volume} {108}},\ \bibinfo
  {pages} {045307} (\bibinfo {year} {2012})}\BibitemShut {NoStop}%
\bibitem [{\citenamefont {Naylor}\ \emph {et~al.}(2016)\citenamefont {Naylor},
  \citenamefont {Brewczyk}, \citenamefont {Gajda}, \citenamefont {Gorceix},
  \citenamefont {Mar\'echal}, \citenamefont {Vernac},\ and\ \citenamefont
  {Laburthe-Tolra}}]{Naylor2016}%
  \BibitemOpen
  \bibfield  {author} {\bibinfo {author} {\bibfnamefont {B.}~\bibnamefont
  {Naylor}}, \bibinfo {author} {\bibfnamefont {M.}~\bibnamefont {Brewczyk}},
  \bibinfo {author} {\bibfnamefont {M.}~\bibnamefont {Gajda}}, \bibinfo
  {author} {\bibfnamefont {O.}~\bibnamefont {Gorceix}}, \bibinfo {author}
  {\bibfnamefont {E.}~\bibnamefont {Mar\'echal}}, \bibinfo {author}
  {\bibfnamefont {L.}~\bibnamefont {Vernac}}, \ and\ \bibinfo {author}
  {\bibfnamefont {B.}~\bibnamefont {Laburthe-Tolra}},\ }\href {\doibase
  10.1103/PhysRevLett.117.185302} {\bibfield  {journal} {\bibinfo  {journal}
  {Phys. Rev. Lett.}\ }\textbf {\bibinfo {volume} {117}},\ \bibinfo {pages}
  {185302} (\bibinfo {year} {2016})}\BibitemShut {NoStop}%
\bibitem [{\citenamefont {Bradley}\ \emph {et~al.}(2000)\citenamefont
  {Bradley}, \citenamefont {McClelland}, \citenamefont {Anderson},\ and\
  \citenamefont {Celotta}}]{Bradley2000}%
  \BibitemOpen
  \bibfield  {author} {\bibinfo {author} {\bibfnamefont {C.~C.}\ \bibnamefont
  {Bradley}}, \bibinfo {author} {\bibfnamefont {J.~J.}\ \bibnamefont
  {McClelland}}, \bibinfo {author} {\bibfnamefont {W.~R.}\ \bibnamefont
  {Anderson}}, \ and\ \bibinfo {author} {\bibfnamefont {R.~J.}\ \bibnamefont
  {Celotta}},\ }\href {\doibase 10.1103/PhysRevA.61.053407} {\bibfield
  {journal} {\bibinfo  {journal} {Phys. Rev. A}\ }\textbf {\bibinfo {volume}
  {61}},\ \bibinfo {pages} {053407} (\bibinfo {year} {2000})}\BibitemShut
  {NoStop}%
\bibitem [{\citenamefont {Chicireanu}\ \emph {et~al.}(2006)\citenamefont
  {Chicireanu}, \citenamefont {Pouderous}, \citenamefont {Barb\'e},
  \citenamefont {Laburthe-Tolra}, \citenamefont {Mar\'echal}, \citenamefont
  {Vernac}, \citenamefont {Keller},\ and\ \citenamefont
  {Gorceix}}]{Chicireanu2006}%
  \BibitemOpen
  \bibfield  {author} {\bibinfo {author} {\bibfnamefont {R.}~\bibnamefont
  {Chicireanu}}, \bibinfo {author} {\bibfnamefont {A.}~\bibnamefont
  {Pouderous}}, \bibinfo {author} {\bibfnamefont {R.}~\bibnamefont {Barb\'e}},
  \bibinfo {author} {\bibfnamefont {B.}~\bibnamefont {Laburthe-Tolra}},
  \bibinfo {author} {\bibfnamefont {E.}~\bibnamefont {Mar\'echal}}, \bibinfo
  {author} {\bibfnamefont {L.}~\bibnamefont {Vernac}}, \bibinfo {author}
  {\bibfnamefont {J.-C.}\ \bibnamefont {Keller}}, \ and\ \bibinfo {author}
  {\bibfnamefont {O.}~\bibnamefont {Gorceix}},\ }\href {\doibase
  10.1103/PhysRevA.73.053406} {\bibfield  {journal} {\bibinfo  {journal} {Phys.
  Rev. A}\ }\textbf {\bibinfo {volume} {73}},\ \bibinfo {pages} {053406}
  (\bibinfo {year} {2006})}\BibitemShut {NoStop}%
\bibitem [{\citenamefont {Schmidt}\ \emph {et~al.}(2003)\citenamefont
  {Schmidt}, \citenamefont {Hensler}, \citenamefont {Werner}, \citenamefont
  {Binhammer}, \citenamefont {G\"orlitz},\ and\ \citenamefont
  {Pfau}}]{Schmidt2003}%
  \BibitemOpen
  \bibfield  {author} {\bibinfo {author} {\bibfnamefont {P.~O.}\ \bibnamefont
  {Schmidt}}, \bibinfo {author} {\bibfnamefont {S.}~\bibnamefont {Hensler}},
  \bibinfo {author} {\bibfnamefont {J.}~\bibnamefont {Werner}}, \bibinfo
  {author} {\bibfnamefont {T.}~\bibnamefont {Binhammer}}, \bibinfo {author}
  {\bibfnamefont {A.}~\bibnamefont {G\"orlitz}}, \ and\ \bibinfo {author}
  {\bibfnamefont {T.}~\bibnamefont {Pfau}},\ }\href {\doibase
  10.1364/JOSAB.20.000960} {\bibfield  {journal} {\bibinfo  {journal} {J. Opt.
  Soc. Am. B}\ }\textbf {\bibinfo {volume} {20}},\ \bibinfo {pages} {960}
  (\bibinfo {year} {2003})}\BibitemShut {NoStop}%
\bibitem [{\citenamefont {Boiron}\ \emph {et~al.}(1995)\citenamefont {Boiron},
  \citenamefont {Trich\'e}, \citenamefont {Meacher}, \citenamefont {Verkerk},\
  and\ \citenamefont {Grynberg}}]{Boiron1995}%
  \BibitemOpen
  \bibfield  {author} {\bibinfo {author} {\bibfnamefont {D.}~\bibnamefont
  {Boiron}}, \bibinfo {author} {\bibfnamefont {C.}~\bibnamefont {Trich\'e}},
  \bibinfo {author} {\bibfnamefont {D.~R.}\ \bibnamefont {Meacher}}, \bibinfo
  {author} {\bibfnamefont {P.}~\bibnamefont {Verkerk}}, \ and\ \bibinfo
  {author} {\bibfnamefont {G.}~\bibnamefont {Grynberg}},\ }\href {\doibase
  10.1103/PhysRevA.52.R3425} {\bibfield  {journal} {\bibinfo  {journal} {Phys.
  Rev. A}\ }\textbf {\bibinfo {volume} {52}},\ \bibinfo {pages} {R3425}
  (\bibinfo {year} {1995})}\BibitemShut {NoStop}%
\bibitem [{\citenamefont {Landini}\ \emph {et~al.}(2011)\citenamefont
  {Landini}, \citenamefont {Roy}, \citenamefont {Carcagn\'{\i}}, \citenamefont
  {Trypogeorgos}, \citenamefont {Fattori}, \citenamefont {Inguscio},\ and\
  \citenamefont {Modugno}}]{Landini2011}%
  \BibitemOpen
  \bibfield  {author} {\bibinfo {author} {\bibfnamefont {M.}~\bibnamefont
  {Landini}}, \bibinfo {author} {\bibfnamefont {S.}~\bibnamefont {Roy}},
  \bibinfo {author} {\bibfnamefont {L.}~\bibnamefont {Carcagn\'{\i}}}, \bibinfo
  {author} {\bibfnamefont {D.}~\bibnamefont {Trypogeorgos}}, \bibinfo {author}
  {\bibfnamefont {M.}~\bibnamefont {Fattori}}, \bibinfo {author} {\bibfnamefont
  {M.}~\bibnamefont {Inguscio}}, \ and\ \bibinfo {author} {\bibfnamefont
  {G.}~\bibnamefont {Modugno}},\ }\href {\doibase 10.1103/PhysRevA.84.043432}
  {\bibfield  {journal} {\bibinfo  {journal} {Phys. Rev. A}\ }\textbf {\bibinfo
  {volume} {84}},\ \bibinfo {pages} {043432} (\bibinfo {year}
  {2011})}\BibitemShut {NoStop}%
\bibitem [{\citenamefont {Rio~Fernandes}\ \emph {et~al.}(2012)\citenamefont
  {Rio~Fernandes}, \citenamefont {Sievers}, \citenamefont {Kretzschmar},
  \citenamefont {Wu}, \citenamefont {Salomon},\ and\ \citenamefont
  {Chevy}}]{Fernandes2012}%
  \BibitemOpen
  \bibfield  {author} {\bibinfo {author} {\bibfnamefont {D.}~\bibnamefont
  {Rio~Fernandes}}, \bibinfo {author} {\bibfnamefont {F.}~\bibnamefont
  {Sievers}}, \bibinfo {author} {\bibfnamefont {N.}~\bibnamefont
  {Kretzschmar}}, \bibinfo {author} {\bibfnamefont {S.}~\bibnamefont {Wu}},
  \bibinfo {author} {\bibfnamefont {C.}~\bibnamefont {Salomon}}, \ and\
  \bibinfo {author} {\bibfnamefont {F.}~\bibnamefont {Chevy}},\ }\href
  {\doibase 10.1209/0295-5075/100/63001} {\bibfield  {journal} {\bibinfo
  {journal} {EPL}\ }\textbf {\bibinfo {volume} {100}},\ \bibinfo {pages}
  {63001} (\bibinfo {year} {2012})}\BibitemShut {NoStop}%
\bibitem [{\citenamefont {Grier}\ \emph {et~al.}(2013)\citenamefont {Grier},
  \citenamefont {Ferrier-Barbut}, \citenamefont {Rem}, \citenamefont
  {Delehaye}, \citenamefont {Khaykovich}, \citenamefont {Chevy},\ and\
  \citenamefont {Salomon}}]{Grier2013}%
  \BibitemOpen
  \bibfield  {author} {\bibinfo {author} {\bibfnamefont {A.~T.}\ \bibnamefont
  {Grier}}, \bibinfo {author} {\bibfnamefont {I.}~\bibnamefont
  {Ferrier-Barbut}}, \bibinfo {author} {\bibfnamefont {B.~S.}\ \bibnamefont
  {Rem}}, \bibinfo {author} {\bibfnamefont {M.}~\bibnamefont {Delehaye}},
  \bibinfo {author} {\bibfnamefont {L.}~\bibnamefont {Khaykovich}}, \bibinfo
  {author} {\bibfnamefont {F.}~\bibnamefont {Chevy}}, \ and\ \bibinfo {author}
  {\bibfnamefont {C.}~\bibnamefont {Salomon}},\ }\href {\doibase
  10.1103/PhysRevA.87.063411} {\bibfield  {journal} {\bibinfo  {journal} {Phys.
  Rev. A}\ }\textbf {\bibinfo {volume} {87}},\ \bibinfo {pages} {063411}
  (\bibinfo {year} {2013})}\BibitemShut {NoStop}%
\bibitem [{\citenamefont {Burchianti}\ \emph {et~al.}(2014)\citenamefont
  {Burchianti}, \citenamefont {Valtolina}, \citenamefont {Seman}, \citenamefont
  {Pace}, \citenamefont {De~Pas}, \citenamefont {Inguscio}, \citenamefont
  {Zaccanti},\ and\ \citenamefont {Roati}}]{Burchianti2014}%
  \BibitemOpen
  \bibfield  {author} {\bibinfo {author} {\bibfnamefont {A.}~\bibnamefont
  {Burchianti}}, \bibinfo {author} {\bibfnamefont {G.}~\bibnamefont
  {Valtolina}}, \bibinfo {author} {\bibfnamefont {J.~A.}\ \bibnamefont
  {Seman}}, \bibinfo {author} {\bibfnamefont {E.}~\bibnamefont {Pace}},
  \bibinfo {author} {\bibfnamefont {M.}~\bibnamefont {De~Pas}}, \bibinfo
  {author} {\bibfnamefont {M.}~\bibnamefont {Inguscio}}, \bibinfo {author}
  {\bibfnamefont {M.}~\bibnamefont {Zaccanti}}, \ and\ \bibinfo {author}
  {\bibfnamefont {G.}~\bibnamefont {Roati}},\ }\href {\doibase
  10.1103/PhysRevA.90.043408} {\bibfield  {journal} {\bibinfo  {journal} {Phys.
  Rev. A}\ }\textbf {\bibinfo {volume} {90}},\ \bibinfo {pages} {043408}
  (\bibinfo {year} {2014})}\BibitemShut {NoStop}%
\bibitem [{\citenamefont {Colzi}\ \emph {et~al.}(2016)\citenamefont {Colzi},
  \citenamefont {Durastante}, \citenamefont {Fava}, \citenamefont {Serafini},
  \citenamefont {Lamporesi},\ and\ \citenamefont {Ferrari}}]{Colzi2016}%
  \BibitemOpen
  \bibfield  {author} {\bibinfo {author} {\bibfnamefont {G.}~\bibnamefont
  {Colzi}}, \bibinfo {author} {\bibfnamefont {G.}~\bibnamefont {Durastante}},
  \bibinfo {author} {\bibfnamefont {E.}~\bibnamefont {Fava}}, \bibinfo {author}
  {\bibfnamefont {S.}~\bibnamefont {Serafini}}, \bibinfo {author}
  {\bibfnamefont {G.}~\bibnamefont {Lamporesi}}, \ and\ \bibinfo {author}
  {\bibfnamefont {G.}~\bibnamefont {Ferrari}},\ }\href {\doibase
  10.1103/PhysRevA.93.023421} {\bibfield  {journal} {\bibinfo  {journal} {Phys.
  Rev. A}\ }\textbf {\bibinfo {volume} {93}},\ \bibinfo {pages} {023421}
  (\bibinfo {year} {2016})}\BibitemShut {NoStop}%
\bibitem [{\citenamefont {Rosi}\ \emph {et~al.}(2018)\citenamefont {Rosi},
  \citenamefont {Burchianti}, \citenamefont {Conclave}, \citenamefont {Naik},
  \citenamefont {Roati}, \citenamefont {Fort},\ and\ \citenamefont
  {Minardi}}]{Rosi2018}%
  \BibitemOpen
  \bibfield  {author} {\bibinfo {author} {\bibfnamefont {S.}~\bibnamefont
  {Rosi}}, \bibinfo {author} {\bibfnamefont {A.}~\bibnamefont {Burchianti}},
  \bibinfo {author} {\bibfnamefont {S.}~\bibnamefont {Conclave}}, \bibinfo
  {author} {\bibfnamefont {D.~S.}\ \bibnamefont {Naik}}, \bibinfo {author}
  {\bibfnamefont {G.}~\bibnamefont {Roati}}, \bibinfo {author} {\bibfnamefont
  {C.}~\bibnamefont {Fort}}, \ and\ \bibinfo {author} {\bibfnamefont
  {F.}~\bibnamefont {Minardi}},\ }\href {\doibase 10.1038/s41598-018-19814-z}
  {\bibfield  {journal} {\bibinfo  {journal} {Sc. Rep.}\ }\textbf {\bibinfo
  {volume} {8}},\ \bibinfo {pages} {1} (\bibinfo {year} {2018})}\BibitemShut
  {NoStop}%
\bibitem [{\citenamefont {Bouton}\ \emph {et~al.}(2015)\citenamefont {Bouton},
  \citenamefont {Chang}, \citenamefont {Hoendervanger}, \citenamefont
  {Nogrette}, \citenamefont {Aspect}, \citenamefont {Westbrook},\ and\
  \citenamefont {Cl\'ement}}]{Bouton2015}%
  \BibitemOpen
  \bibfield  {author} {\bibinfo {author} {\bibfnamefont {Q.}~\bibnamefont
  {Bouton}}, \bibinfo {author} {\bibfnamefont {R.}~\bibnamefont {Chang}},
  \bibinfo {author} {\bibfnamefont {A.~L.}\ \bibnamefont {Hoendervanger}},
  \bibinfo {author} {\bibfnamefont {F.}~\bibnamefont {Nogrette}}, \bibinfo
  {author} {\bibfnamefont {A.}~\bibnamefont {Aspect}}, \bibinfo {author}
  {\bibfnamefont {C.~I.}\ \bibnamefont {Westbrook}}, \ and\ \bibinfo {author}
  {\bibfnamefont {D.}~\bibnamefont {Cl\'ement}},\ }\href {\doibase
  10.1103/PhysRevA.91.061402} {\bibfield  {journal} {\bibinfo  {journal} {Phys.
  Rev. A}\ }\textbf {\bibinfo {volume} {91}},\ \bibinfo {pages} {061402}
  (\bibinfo {year} {2015})}\BibitemShut {NoStop}%
\bibitem [{\citenamefont {Drewsen}\ \emph {et~al.}(1996)\citenamefont
  {Drewsen}, \citenamefont {Drodofsky}, \citenamefont {Weber}, \citenamefont
  {Schreiber},\ and\ \citenamefont {Mlynek}}]{Drewsen1996}%
  \BibitemOpen
  \bibfield  {author} {\bibinfo {author} {\bibfnamefont {M.}~\bibnamefont
  {Drewsen}}, \bibinfo {author} {\bibfnamefont {U.}~\bibnamefont {Drodofsky}},
  \bibinfo {author} {\bibfnamefont {C.}~\bibnamefont {Weber}}, \bibinfo
  {author} {\bibfnamefont {G.}~\bibnamefont {Schreiber}}, \ and\ \bibinfo
  {author} {\bibfnamefont {J.}~\bibnamefont {Mlynek}},\ }\href {\doibase
  10.1088/0953-4075/29/23/006} {\bibfield  {journal} {\bibinfo  {journal} {J.
  Phys. B: At. Mol. Opt. Phys.}\ }\textbf {\bibinfo {volume} {29}},\ \bibinfo
  {pages} {L843} (\bibinfo {year} {1996})}\BibitemShut {NoStop}%
\bibitem [{\citenamefont {Hopkins}\ and\ \citenamefont
  {Durrant}(1997)}]{Hopkins1997}%
  \BibitemOpen
  \bibfield  {author} {\bibinfo {author} {\bibfnamefont {S.~A.}\ \bibnamefont
  {Hopkins}}\ and\ \bibinfo {author} {\bibfnamefont {A.~V.}\ \bibnamefont
  {Durrant}},\ }\href {\doibase 10.1103/PhysRevA.56.4012} {\bibfield  {journal}
  {\bibinfo  {journal} {Phys. Rev. A}\ }\textbf {\bibinfo {volume} {56}},\
  \bibinfo {pages} {4012} (\bibinfo {year} {1997})}\BibitemShut {NoStop}%
\bibitem [{\citenamefont {Weidem\"uller}\ \emph {et~al.}(1994)\citenamefont
  {Weidem\"uller}, \citenamefont {Esslinger}, \citenamefont {Ol'shanii},
  \citenamefont {Hemmerich},\ and\ \citenamefont {H\"ansch}}]{Weidemuller1994}%
  \BibitemOpen
  \bibfield  {author} {\bibinfo {author} {\bibfnamefont {M.}~\bibnamefont
  {Weidem\"uller}}, \bibinfo {author} {\bibfnamefont {T.}~\bibnamefont
  {Esslinger}}, \bibinfo {author} {\bibfnamefont {M.~A.}\ \bibnamefont
  {Ol'shanii}}, \bibinfo {author} {\bibfnamefont {A.}~\bibnamefont
  {Hemmerich}}, \ and\ \bibinfo {author} {\bibfnamefont {T.~W.}\ \bibnamefont
  {H\"ansch}},\ }\href {\doibase 10.1209/0295-5075} {\bibfield  {journal}
  {\bibinfo  {journal} {EPL}\ }\textbf {\bibinfo {volume} {27}},\ \bibinfo
  {pages} {109} (\bibinfo {year} {1994})}\BibitemShut {NoStop}%
\bibitem [{\citenamefont {Dalibard}\ and\ \citenamefont
  {Cohen-Tannoudji}(1989)}]{Dalibard1989}%
  \BibitemOpen
  \bibfield  {author} {\bibinfo {author} {\bibfnamefont {J.}~\bibnamefont
  {Dalibard}}\ and\ \bibinfo {author} {\bibfnamefont {C.}~\bibnamefont
  {Cohen-Tannoudji}},\ }\href {\doibase 10.1364/JOSAB.6.002023} {\bibfield
  {journal} {\bibinfo  {journal} {J. Opt. Soc. Am. B}\ }\textbf {\bibinfo
  {volume} {6}},\ \bibinfo {pages} {2023} (\bibinfo {year} {1989})}\BibitemShut
  {NoStop}%
\bibitem [{\citenamefont {Boiron}\ \emph {et~al.}(1998)\citenamefont {Boiron},
  \citenamefont {Michaud}, \citenamefont {Fournier}, \citenamefont {Simard},
  \citenamefont {Sprenger}, \citenamefont {Grynberg},\ and\ \citenamefont
  {Salomon}}]{Boiron1998}%
  \BibitemOpen
  \bibfield  {author} {\bibinfo {author} {\bibfnamefont {D.}~\bibnamefont
  {Boiron}}, \bibinfo {author} {\bibfnamefont {A.}~\bibnamefont {Michaud}},
  \bibinfo {author} {\bibfnamefont {J.~M.}\ \bibnamefont {Fournier}}, \bibinfo
  {author} {\bibfnamefont {L.}~\bibnamefont {Simard}}, \bibinfo {author}
  {\bibfnamefont {M.}~\bibnamefont {Sprenger}}, \bibinfo {author}
  {\bibfnamefont {G.}~\bibnamefont {Grynberg}}, \ and\ \bibinfo {author}
  {\bibfnamefont {C.}~\bibnamefont {Salomon}},\ }\href {\doibase
  10.1103/PhysRevA.57.R4106} {\bibfield  {journal} {\bibinfo  {journal} {Phys.
  Rev. A}\ }\textbf {\bibinfo {volume} {57}},\ \bibinfo {pages} {R4106}
  (\bibinfo {year} {1998})}\BibitemShut {NoStop}%
\bibitem [{\citenamefont {DePue}\ \emph {et~al.}(1999)\citenamefont {DePue},
  \citenamefont {McCormick}, \citenamefont {Winoto}, \citenamefont {Oliver},\
  and\ \citenamefont {Weiss}}]{DePue1999}%
  \BibitemOpen
  \bibfield  {author} {\bibinfo {author} {\bibfnamefont {M.~T.}\ \bibnamefont
  {DePue}}, \bibinfo {author} {\bibfnamefont {C.}~\bibnamefont {McCormick}},
  \bibinfo {author} {\bibfnamefont {S.~L.}\ \bibnamefont {Winoto}}, \bibinfo
  {author} {\bibfnamefont {S.}~\bibnamefont {Oliver}}, \ and\ \bibinfo {author}
  {\bibfnamefont {D.~S.}\ \bibnamefont {Weiss}},\ }\href {\doibase
  10.1103/PhysRevLett.82.2262} {\bibfield  {journal} {\bibinfo  {journal}
  {Phys. Rev. Lett.}\ }\textbf {\bibinfo {volume} {82}},\ \bibinfo {pages}
  {2262} (\bibinfo {year} {1999})}\BibitemShut {NoStop}%
\bibitem [{\citenamefont {Hu}\ \emph {et~al.}(2017)\citenamefont {Hu},
  \citenamefont {Urvoy}, \citenamefont {Vendeiro}, \citenamefont {Cr\'epel},
  \citenamefont {Chen},\ and\ \citenamefont {Vuleti\'c}}]{Hu2017}%
  \BibitemOpen
  \bibfield  {author} {\bibinfo {author} {\bibfnamefont {J.}~\bibnamefont
  {Hu}}, \bibinfo {author} {\bibfnamefont {A.}~\bibnamefont {Urvoy}}, \bibinfo
  {author} {\bibfnamefont {Z.}~\bibnamefont {Vendeiro}}, \bibinfo {author}
  {\bibfnamefont {V.}~\bibnamefont {Cr\'epel}}, \bibinfo {author}
  {\bibfnamefont {W.}~\bibnamefont {Chen}}, \ and\ \bibinfo {author}
  {\bibfnamefont {V.}~\bibnamefont {Vuleti\'c}},\ }\href {\doibase
  10.1126/science.aan5614} {\bibfield  {journal} {\bibinfo  {journal}
  {Science}\ }\textbf {\bibinfo {volume} {358}},\ \bibinfo {pages} {1078}
  (\bibinfo {year} {2017})}\BibitemShut {NoStop}%
\bibitem [{\citenamefont {Ammann}\ and\ \citenamefont
  {Christensen}(1997)}]{Ammann1997}%
  \BibitemOpen
  \bibfield  {author} {\bibinfo {author} {\bibfnamefont {H.}~\bibnamefont
  {Ammann}}\ and\ \bibinfo {author} {\bibfnamefont {N.}~\bibnamefont
  {Christensen}},\ }\href {\doibase 10.1103/PhysRevLett.78.2088} {\bibfield
  {journal} {\bibinfo  {journal} {Phys. Rev. Lett.}\ }\textbf {\bibinfo
  {volume} {78}},\ \bibinfo {pages} {2088} (\bibinfo {year}
  {1997})}\BibitemShut {NoStop}%
\bibitem [{\citenamefont {Mar\'echal}\ \emph {et~al.}(1999)\citenamefont
  {Mar\'echal}, \citenamefont {Guibal}, \citenamefont {Bossennec},
  \citenamefont {Barb\'e}, \citenamefont {Keller},\ and\ \citenamefont
  {Gorceix}}]{Marechal1999}%
  \BibitemOpen
  \bibfield  {author} {\bibinfo {author} {\bibfnamefont {E.}~\bibnamefont
  {Mar\'echal}}, \bibinfo {author} {\bibfnamefont {S.}~\bibnamefont {Guibal}},
  \bibinfo {author} {\bibfnamefont {J.-L.}\ \bibnamefont {Bossennec}}, \bibinfo
  {author} {\bibfnamefont {R.}~\bibnamefont {Barb\'e}}, \bibinfo {author}
  {\bibfnamefont {J.-C.}\ \bibnamefont {Keller}}, \ and\ \bibinfo {author}
  {\bibfnamefont {O.}~\bibnamefont {Gorceix}},\ }\href {\doibase
  10.1103/PhysRevA.59.4636} {\bibfield  {journal} {\bibinfo  {journal} {Phys.
  Rev. A}\ }\textbf {\bibinfo {volume} {59}},\ \bibinfo {pages} {4636}
  (\bibinfo {year} {1999})}\BibitemShut {NoStop}%
\bibitem [{\citenamefont {Bismut}\ \emph {et~al.}(2011)\citenamefont {Bismut},
  \citenamefont {Pasquiou}, \citenamefont {Ciampini}, \citenamefont
  {Laburthe-Tolra}, \citenamefont {Mar\'echal}, \citenamefont {Vernac},\ and\
  \citenamefont {Gorceix}}]{Bismut2011}%
  \BibitemOpen
  \bibfield  {author} {\bibinfo {author} {\bibfnamefont {G.}~\bibnamefont
  {Bismut}}, \bibinfo {author} {\bibfnamefont {B.}~\bibnamefont {Pasquiou}},
  \bibinfo {author} {\bibfnamefont {D.}~\bibnamefont {Ciampini}}, \bibinfo
  {author} {\bibfnamefont {B.}~\bibnamefont {Laburthe-Tolra}}, \bibinfo
  {author} {\bibfnamefont {E.}~\bibnamefont {Mar\'echal}}, \bibinfo {author}
  {\bibfnamefont {L.}~\bibnamefont {Vernac}}, \ and\ \bibinfo {author}
  {\bibfnamefont {O.}~\bibnamefont {Gorceix}},\ }\href {\doibase
  10.1007/s00340-010-4171-y} {\bibfield  {journal} {\bibinfo  {journal} {App.
  Phys. B}\ }\textbf {\bibinfo {volume} {102}},\ \bibinfo {pages} {1} (\bibinfo
  {year} {2011})}\BibitemShut {NoStop}%
\bibitem [{\citenamefont {Beaufils}\ \emph
  {et~al.}(2008{\natexlab{b}})\citenamefont {Beaufils}, \citenamefont
  {Chicireanu}, \citenamefont {Pouderous}, \citenamefont {de~Souza~Melo},
  \citenamefont {Laburthe-Tolra}, \citenamefont {Mar\'echal}, \citenamefont
  {Vernac}, \citenamefont {Keller},\ and\ \citenamefont
  {Gorceix}}]{Beaufils2008RF}%
  \BibitemOpen
  \bibfield  {author} {\bibinfo {author} {\bibfnamefont {Q.}~\bibnamefont
  {Beaufils}}, \bibinfo {author} {\bibfnamefont {R.}~\bibnamefont
  {Chicireanu}}, \bibinfo {author} {\bibfnamefont {A.}~\bibnamefont
  {Pouderous}}, \bibinfo {author} {\bibfnamefont {W.}~\bibnamefont
  {de~Souza~Melo}}, \bibinfo {author} {\bibfnamefont {B.}~\bibnamefont
  {Laburthe-Tolra}}, \bibinfo {author} {\bibfnamefont {E.}~\bibnamefont
  {Mar\'echal}}, \bibinfo {author} {\bibfnamefont {L.}~\bibnamefont {Vernac}},
  \bibinfo {author} {\bibfnamefont {J.~C.}\ \bibnamefont {Keller}}, \ and\
  \bibinfo {author} {\bibfnamefont {O.}~\bibnamefont {Gorceix}},\ }\href
  {\doibase 10.1103/PhysRevA.77.053413} {\bibfield  {journal} {\bibinfo
  {journal} {Phys. Rev. A}\ }\textbf {\bibinfo {volume} {77}},\ \bibinfo
  {pages} {053413} (\bibinfo {year} {2008}{\natexlab{b}})}\BibitemShut
  {NoStop}%
\bibitem [{\citenamefont {Chicireanu}\ \emph {et~al.}(2007)\citenamefont
  {Chicireanu}, \citenamefont {Beaufils}, \citenamefont {Pouderous},
  \citenamefont {Laburthe-Tolra}, \citenamefont {Mar\'echal}, \citenamefont
  {Vernac}, \citenamefont {Keller},\ and\ \citenamefont
  {Gorceix}}]{Chicireanu2007}%
  \BibitemOpen
  \bibfield  {author} {\bibinfo {author} {\bibfnamefont {R.}~\bibnamefont
  {Chicireanu}}, \bibinfo {author} {\bibfnamefont {Q.}~\bibnamefont
  {Beaufils}}, \bibinfo {author} {\bibfnamefont {A.}~\bibnamefont {Pouderous}},
  \bibinfo {author} {\bibfnamefont {B.}~\bibnamefont {Laburthe-Tolra}},
  \bibinfo {author} {\bibfnamefont {E.}~\bibnamefont {Mar\'echal}}, \bibinfo
  {author} {\bibfnamefont {L.}~\bibnamefont {Vernac}}, \bibinfo {author}
  {\bibfnamefont {J.-C.}\ \bibnamefont {Keller}}, \ and\ \bibinfo {author}
  {\bibfnamefont {O.}~\bibnamefont {Gorceix}},\ }\href {\doibase
  10.1140/epjd/e2007-00245-y} {\bibfield  {journal} {\bibinfo  {journal} {Eur.
  Phys. J. D}\ }\textbf {\bibinfo {volume} {45}},\ \bibinfo {pages} {189}
  (\bibinfo {year} {2007})}\BibitemShut {NoStop}%
\bibitem [{\citenamefont {Volchkov}\ \emph {et~al.}(2014)\citenamefont
  {Volchkov}, \citenamefont {R\"uhrig}, \citenamefont {Pfau},\ and\
  \citenamefont {Griesmaier}}]{Volochkov2014}%
  \BibitemOpen
  \bibfield  {author} {\bibinfo {author} {\bibfnamefont {V.~V.}\ \bibnamefont
  {Volchkov}}, \bibinfo {author} {\bibfnamefont {J.}~\bibnamefont {R\"uhrig}},
  \bibinfo {author} {\bibfnamefont {T.}~\bibnamefont {Pfau}}, \ and\ \bibinfo
  {author} {\bibfnamefont {A.}~\bibnamefont {Griesmaier}},\ }\href {\doibase
  10.1103/PhysRevA.89.043417} {\bibfield  {journal} {\bibinfo  {journal} {Phys.
  Rev. A}\ }\textbf {\bibinfo {volume} {89}},\ \bibinfo {pages} {043417}
  (\bibinfo {year} {2014})}\BibitemShut {NoStop}%
\bibitem [{\citenamefont {Trich{\'e}}\ \emph {et~al.}(1999)\citenamefont
  {Trich{\'e}}, \citenamefont {Verkerk},\ and\ \citenamefont
  {Grynberg}}]{Triche1999}%
  \BibitemOpen
  \bibfield  {author} {\bibinfo {author} {\bibfnamefont {C.}~\bibnamefont
  {Trich{\'e}}}, \bibinfo {author} {\bibfnamefont {P.}~\bibnamefont {Verkerk}},
  \ and\ \bibinfo {author} {\bibfnamefont {G.}~\bibnamefont {Grynberg}},\
  }\href {\doibase 10.1007/s100530050249} {\bibfield  {journal} {\bibinfo
  {journal} {Eur. Phys. J. D}\ }\textbf {\bibinfo {volume} {5}},\ \bibinfo
  {pages} {225} (\bibinfo {year} {1999})}\BibitemShut {NoStop}%
\bibitem [{\citenamefont {Nath}\ \emph {et~al.}(2013)\citenamefont {Nath},
  \citenamefont {Easwaran}, \citenamefont {Rajalakshmi},\ and\ \citenamefont
  {Unnikrishnan}}]{Nath2013}%
  \BibitemOpen
  \bibfield  {author} {\bibinfo {author} {\bibfnamefont {D.}~\bibnamefont
  {Nath}}, \bibinfo {author} {\bibfnamefont {R.~K.}\ \bibnamefont {Easwaran}},
  \bibinfo {author} {\bibfnamefont {G.}~\bibnamefont {Rajalakshmi}}, \ and\
  \bibinfo {author} {\bibfnamefont {C.~S.}\ \bibnamefont {Unnikrishnan}},\
  }\href {\doibase 10.1103/PhysRevA.88.053407} {\bibfield  {journal} {\bibinfo
  {journal} {Phys. Rev. A}\ }\textbf {\bibinfo {volume} {88}},\ \bibinfo
  {pages} {053407} (\bibinfo {year} {2013})}\BibitemShut {NoStop}%
\bibitem [{\citenamefont {Salomon}\ \emph {et~al.}(2013)\citenamefont
  {Salomon}, \citenamefont {Fouché}, \citenamefont {Wang}, \citenamefont
  {Aspect}, \citenamefont {Bouyer},\ and\ \citenamefont
  {Bourdel}}]{Salomon2013}%
  \BibitemOpen
  \bibfield  {author} {\bibinfo {author} {\bibfnamefont {G.}~\bibnamefont
  {Salomon}}, \bibinfo {author} {\bibfnamefont {L.}~\bibnamefont {Fouché}},
  \bibinfo {author} {\bibfnamefont {P.}~\bibnamefont {Wang}}, \bibinfo {author}
  {\bibfnamefont {A.}~\bibnamefont {Aspect}}, \bibinfo {author} {\bibfnamefont
  {P.}~\bibnamefont {Bouyer}}, \ and\ \bibinfo {author} {\bibfnamefont
  {T.}~\bibnamefont {Bourdel}},\ }\href
  {http://stacks.iop.org/0295-5075/104/i=6/a=63002} {\bibfield  {journal}
  {\bibinfo  {journal} {EPL}\ }\textbf {\bibinfo {volume} {104}},\ \bibinfo
  {pages} {63002} (\bibinfo {year} {2013})}\BibitemShut {NoStop}%
\bibitem [{\citenamefont {Sievers}\ \emph {et~al.}(2015)\citenamefont
  {Sievers}, \citenamefont {Kretzschmar}, \citenamefont {Fernandes},
  \citenamefont {Suchet}, \citenamefont {Rabinovic}, \citenamefont {Wu},
  \citenamefont {Parker}, \citenamefont {Khaykovich}, \citenamefont {Salomon},\
  and\ \citenamefont {Chevy}}]{Sievers2015}%
  \BibitemOpen
  \bibfield  {author} {\bibinfo {author} {\bibfnamefont {F.}~\bibnamefont
  {Sievers}}, \bibinfo {author} {\bibfnamefont {N.}~\bibnamefont
  {Kretzschmar}}, \bibinfo {author} {\bibfnamefont {D.~R.}\ \bibnamefont
  {Fernandes}}, \bibinfo {author} {\bibfnamefont {D.}~\bibnamefont {Suchet}},
  \bibinfo {author} {\bibfnamefont {M.}~\bibnamefont {Rabinovic}}, \bibinfo
  {author} {\bibfnamefont {S.}~\bibnamefont {Wu}}, \bibinfo {author}
  {\bibfnamefont {C.~V.}\ \bibnamefont {Parker}}, \bibinfo {author}
  {\bibfnamefont {L.}~\bibnamefont {Khaykovich}}, \bibinfo {author}
  {\bibfnamefont {C.}~\bibnamefont {Salomon}}, \ and\ \bibinfo {author}
  {\bibfnamefont {F.}~\bibnamefont {Chevy}},\ }\href {\doibase
  10.1103/PhysRevA.91.023426} {\bibfield  {journal} {\bibinfo  {journal} {Phys.
  Rev. A}\ }\textbf {\bibinfo {volume} {91}},\ \bibinfo {pages} {023426}
  (\bibinfo {year} {2015})}\BibitemShut {NoStop}%
\bibitem [{\citenamefont {Tarnowski}(2015)}]{Tarnowski2015}%
  \BibitemOpen
  \bibfield  {author} {\bibinfo {author} {\bibfnamefont {M.}~\bibnamefont
  {Tarnowski}},\ }\emph {\bibinfo {title} {Implementation and Characterization
  of a Gray Molasses and of Tunable Hexagonal Optical Lattices for
  ${}^{40}$K}},\ \href
  {http://photon.physnet.uni-hamburg.de/fileadmin/user_upload/ILP/Sengstock/Research/BFM/Theses/Masterthesis-Matthias-Tarnowski.pdf}
  {Master's thesis},\ \bibinfo  {school} {Universität Hamburg} (\bibinfo
  {year} {2015})\BibitemShut {NoStop}%
\bibitem [{\citenamefont {Chen}\ \emph {et~al.}(2016)\citenamefont {Chen},
  \citenamefont {Yao}, \citenamefont {Wu}, \citenamefont {Liu}, \citenamefont
  {Wang}, \citenamefont {Wang}, \citenamefont {Chen},\ and\ \citenamefont
  {Pan}}]{Chen2016}%
  \BibitemOpen
  \bibfield  {author} {\bibinfo {author} {\bibfnamefont {H.-Z.}\ \bibnamefont
  {Chen}}, \bibinfo {author} {\bibfnamefont {X.-C.}\ \bibnamefont {Yao}},
  \bibinfo {author} {\bibfnamefont {Y.-P.}\ \bibnamefont {Wu}}, \bibinfo
  {author} {\bibfnamefont {X.-P.}\ \bibnamefont {Liu}}, \bibinfo {author}
  {\bibfnamefont {X.-Q.}\ \bibnamefont {Wang}}, \bibinfo {author}
  {\bibfnamefont {Y.-X.}\ \bibnamefont {Wang}}, \bibinfo {author}
  {\bibfnamefont {Y.-A.}\ \bibnamefont {Chen}}, \ and\ \bibinfo {author}
  {\bibfnamefont {J.-W.}\ \bibnamefont {Pan}},\ }\href {\doibase
  10.1103/PhysRevA.94.033408} {\bibfield  {journal} {\bibinfo  {journal} {Phys.
  Rev. A}\ }\textbf {\bibinfo {volume} {94}},\ \bibinfo {pages} {033408}
  (\bibinfo {year} {2016})}\BibitemShut {NoStop}%
\bibitem [{\citenamefont {Kramida}\ \emph {et~al.}(2018)\citenamefont
  {Kramida}, \citenamefont {Ralchenko}, \citenamefont {Reader},\ and\
  \citenamefont {{ NIST ASD Team}}}]{NIST_ASD}%
  \BibitemOpen
  \bibfield  {author} {\bibinfo {author} {\bibfnamefont {A.}~\bibnamefont
  {Kramida}}, \bibinfo {author} {\bibfnamefont {Y.}~\bibnamefont {Ralchenko}},
  \bibinfo {author} {\bibfnamefont {J.}~\bibnamefont {Reader}}, \ and\ \bibinfo
  {author} {\bibnamefont {{ NIST ASD Team}}},\ }\href {\doibase
  10.18434/T4W30F} {\enquote {\bibinfo {title} {{NIST Atomic Spectra Database
  (ver. 5.6.1)}},}\ } (\bibinfo {year} {2018})\BibitemShut {NoStop}%
\bibitem [{\citenamefont {Steck}\ \emph {et~al.}()\citenamefont {Steck},
  \citenamefont {Gehm},\ and\ \citenamefont {Tiecke}}]{Alkali_data}%
  \BibitemOpen
  \bibfield  {author} {\bibinfo {author} {\bibfnamefont {D.}~\bibnamefont
  {Steck}}, \bibinfo {author} {\bibfnamefont {M.}~\bibnamefont {Gehm}}, \ and\
  \bibinfo {author} {\bibfnamefont {T.}~\bibnamefont {Tiecke}},\ }\href
  {https://steck.us/alkalidata/} {\enquote {\bibinfo {title} {Alkali atoms
  properties},}\ }\BibitemShut {NoStop}%
\end{thebibliography}%

\end{document}